\documentclass[11pt,a4paper,oneside]{article}
\usepackage{amssymb}
\usepackage{amsmath}
\usepackage{framed}
\usepackage{slashed}
\usepackage{graphicx}
\usepackage{indentfirst}
\usepackage{a4wide}
\usepackage[utf8]{inputenc}
\usepackage[small,bf,normal]{caption}
\usepackage{xspace}
\usepackage{hyperref}
\usepackage{latexsym}
\usepackage{units}
\usepackage{bbold}
\usepackage{braket}
\usepackage{cancel}
\usepackage{color}
\usepackage[normalem]{ulem}
\usepackage{relsize}
\usepackage{adjustbox}
\usepackage{import}
\usepackage{pdfsync}
\usepackage{bm}
\usepackage{array}
\usepackage{tikz}
\usetikzlibrary{math}
\usepackage{epstopdf}
\usepackage{mathtools}
\usepackage{siunitx}
\usepackage{subfig}
\usepackage{breakcites}
\usepackage{etoolbox}
\usepackage{natbib}

\renewcommand{\vec}[1]{\bm{#1}}
\renewcommand{\d}{\mathrm{d}}

\makeatletter

\newcommand*{\myfnsymbolsingle}[1]{%
  \ensuremath{%
    \ifcase#1% 0
    \or % 2
    \dagger
    \or % 3  
    \ddagger
    \or % 4   
    \mathsection
    \or % 5
    \mathparagraph
    \or
    \#
    \else % >= 6
    \@ctrerr  
    \fi
  }%   
}   
\makeatother

\newcommand{\hcp}{\text{HCP}}
\newcommand{\prp}{\text{PRP}}
\newcommand{\prpv}{\text{PRP,V}}
\newcommand{\prph}{\text{PRP,H}}
\newcommand{\vcp}{\text{VCP}}

\usepackage{alphalph}
\newalphalph{\myfnsymbolmult}[mult]{\myfnsymbolsingle}{}

\title{\textbf{A damped forward EMI model for a horizontally stratified earth}}
\author{{Steven Delrue~$^{a}$\thanks{steven.delrue@kuleuven.be}, David Dudal~$^{a,b}$\thanks{david.dudal@kuleuven.be}, Benjamin Maveau~$^{a}$\thanks{benjamin.maveau@kuleuven.be (corresponding author); +32 56 24 64 71}}\\\\
\textit{{\small $^a$ KU Leuven Campus Kulak Kortrijk -- Kulak, Department of Physics, Etienne Sabbelaan 53 box 7657,}}\\
\textit{{\small 8500 Kortrijk, Belgium}}\\
\textit{{\small $^b$           Ghent University, Department of Physics and Astronomy, Krijgslaan 281-S9, 9000 Gent, Belgium}}}

\begin{document}
\maketitle
\clearpage
\begin{abstract}
  \noindent If a magnetic dipole is placed above the surface of the earth, the Electromagnetic Induction (EMI) effect, encoded in Maxwell's equations, causes eddy currents in the soil which, on their turn, induce response electromagnetic fields. The magnetic field can be measured in geophysical surveys to determine the conductivity profile of the ground in a non-destructive manner. The forward model used in the inversion of  experimental data usually consists of a set of horizontal homogeneous layers. A frequently used analytical model, proposed by McNeill, does not include the interaction between the eddy currents, and therefore fails for larger conductivities. In this paper we construct a new forward, analytical, model to estimate the magnetic field caused by a horizontally stratified earth but which approximates the interaction between eddy currents. This makes it valid for a broader range of parameters than the current state of the art. Furthermore, the error with the (numerically obtainable) exact result is substantially decreased. We also calculate the vertical sensitivity (``depth of exploration'') of the model and observe that it is in good agreement with the values obtained from the exact model.\\
  EMI: electromagnetic induction; LIN: low induction number; HCP: horizontal coplanar; PRP: perpendicular
\end{abstract}
\section{Introduction}
\label{sec:introduction}
\setcounter{footnote}{1}
From EMI surveys one can reconstruct, using an appropriate inversion algorithm, an approximate conductivity profile of the soil. Such profile can, for example, be used to measure the soil salinity~\citep{hendrickx_soil_1992}, detect anomalies~\citep{de_smedt_unveiling_2014,bongiovanni_rapid_2008}, monitor soil contamination~\citep{senos_matias_geophysical_1994}, non-invasively prospect for archeological features~\citep{saey_electrical_2012} or probe for salty seawater intrusion into groundwater reservoirs~\citep{holman_land_1998,himi_geophysical_2017,moghadas_joint_2010}. A successful inversion requires a forward model which approximates the exact result sufficiently accurately, but at the same time allows for a stable and relatively fast numerical solution.

A common model used in EMI surveys is the approach \cite{mcneill_electromagnetic_1980} proposed based on the work of \cite{wait_induction_1954,wait_note_1962}. Slicing the subsurface into an infinite number of very thin sheets, one calculates the contribution of one such sheet due to the varying magnetic dipole. Summing all these contributions results in the total magnetic response field, from now on called the secondary field. When operating at Low Induction Number (LIN) the obtained solution approximates the exact solution relatively well.

The LIN assumption fails  when the frequency of the dipole $f$, the electrical conductivity $\sigma$ and/or the distance between emitter and receivers $s$ are large enough so that $2\pi f\mu\sigma s^{2}$ is much larger than 1. A high conductivity occurs for measurements of saline soil while the a larger intercoil spacing is used to characterise the deeper parts of the soil. Indeed, a larger intercoil distance increases the contribution of the lower regions, causing a larger influence in the secondary field. The effects of high saline grounds have been studied in e.g.~\cite{delefortrie_frequency_2014}.~\cite{reid_doubling_1999} discussed the effect of the conductivity on the attenuation and concluded that it, together with the intercoil distance and frequency, strongly affects the decay of the electromagnetic fields. Therefore a more complete model is required to describe highly conductive layers.

Despite these limitations, the data collected based on the LIN assumption are able to obtain a good estimate of the conductivity profile~\citep{hendrickx_inversion_2002,saey_comparing_2015} under the right circumstances. A huge advantage of the McNeill reduction is the linearity in the conductivity and the simplicity of the equations. These features make it an excellent model for the initialization of an inversion scheme~\citep{mester_quantitative_2011}.

An alternative approach is to determine the exact solution in case of a layered earth. \cite{wait_geo-electromagnetism_1982} and \cite{frischknecht_electrical_1966}, derived for this configuration a recursion relation allowing one to calculate the secondary field directly. Despite some promising results (e.g.~\citep{hendrickx_inversion_2002,mester_quantitative_2011,minsley_trans-dimensional_2011,triantafilis_modeling_2012,saey_comparing_2015}) obtained using these models in recent years, there are still several reasons justifying the development of an approximate analytic (forward) solution, as intended in the current paper:
  \begin{itemize}
\item The basic LIN model is still widely used~\footnote{See for example the citation list to the technical note of McNeill using Google Scholar.}. We expect this is related to its inherent simplicity and, perhaps, also partially caused by the fact that the commonly used instruments, like the DUALEM and those from GEONICS, effectively generate their data in terms of the apparent conductivity, as introduced in the McNeill derivation. A procedure by \cite{beamish_low_2011} suggests an instrument-dependent relation between the measured apparent conductivity and the LIN-equivalent apparent conductivity, allowing one to map the soil for all induction numbers. However this procedure cannot be applied for inversion and, the intrinsic shortcomings of the LIN approximation remain and were recently restated in~\cite{hatch_environmental_2017}, as well as in \cite{reid_application_2001}. 
\item The choice for commercial or free software packages implies the user has to rely on the preprogrammed inversion strategy. For example, according to~\cite{constable_occams_1987} and following~\cite{triantafilis_modeling_2012}, {\sf EM4Soil} is based on a relatively simple Tikhonov regularization with $L_2$ norm~\citep{kirsch_introduction_2011}, which is known to enforce smoothness of the estimated inverse solution. It is easy to imagine that this at times can be a rather undesirable feature, e.g.~in the presence of blocky structures.  Therefore, in the course of this paper and a forthcoming follow-up study of the inverse problem, we prefer not to rely on existing software. 
\item The error calculated in our proposed analytical model falls within the typical accuracy of most instruments (e.g.~\cite{geonics_limited_geophysical_2018} and~\cite{gf_instruments_gf_2018}). Hence, in practice, our approximate solution is as powerful as any more elaborate numerical scheme.
\item Finally, the proposed new model can serve as a stepping stone to a model for surveys in seawater. Another possibility is the extrapolation of our model to 2D (or even 3D).
\end{itemize}

In all derivations, we assume that the relative magnetic permeability is always equal to one. The displacement currents are neglected due to the low frequency and short intercoil spacing. All derivations are performed in the frequency domain, therefore the notation is simplified by omitting the complex exponential factor $(\exp{(iwt)})$ in all physical fields. This corresponds to the quasi-stationary field regime.

\section{Survey of the iterative solution for an $N$-layer model}
\label{sec:solution-n-layer}

\begin{figure}[t]
  \centering
  \begin{tikzpicture}[>=stealth]
    \def\m{.5}
    \def\w{3}
    \def\h{1,1.5,2.5,3,5}

    \draw[->] (0,0) -- (0,\m) node [midway,right] {$\vec{m}$};
    \draw[<->] (\w,0) -- (\w, -1) node [midway,right] {$h$};

    \foreach \hi [count=\i from 0,remember=\hi as \prevhi (initially 0)] in \h {
      \draw (-\w,-\hi) -- (\w,-\hi);

      \node at ({-\w*0.5}, {-0.5*\hi-0.5*\prevhi}) {$\sigma_{\i}$};
      \ifthenelse{\NOT\i=0}{
        \node at ({\w*.5},{-0.5*\hi-0.5*\prevhi}) {Soil layer \i};
        \draw[<->] (\w,-\prevhi) -- (\w, -\hi) node [midway,right] {$\Delta h_{\i}$};
      } {
        \node at ({\w*.5},{-0.5*\hi-0.5*\prevhi}) {Air};
      }
    }

    \draw[<->] (-\w-.1,0) -- (-\w-.1,-3) node [midway, left] {$h_{3}$};
    \node at (0,-5.5) {$\vdots$};
  \end{tikzpicture}
  \caption{An axial symmetric problem consisting of a half-space of air and $N$ layers of soil each with a variable conductivity ($\sigma_{i}$) and thickness ($\Delta h_{i}$).}\label{fig:n-layer}
\end{figure}

When a vertical~\footnote{A derivation for a horizontal dipole is given in Appendix~\ref{sec:horizontal-dipole}.} magnetic dipole is placed a height $h$ above a horizontally stratified earth, we can reduce the problem to an axial-symmetric system consisting of $N$ layers each with a different conductivity $\sigma_i$, as illustrated in Figure~\ref{fig:n-layer}. The Maxwell's equations in the frequency domain are~\citep{jackson_classical_1975}:
\begin{align}
  \label{eq:10}
  \vec{\nabla}\cdot\vec{E} &= \frac{\rho}{\epsilon_{0}}= 0, & \vec{\nabla}\cdot\vec{H} &= 0,\\
  \vec{\nabla}\times\vec{E} &= -i\mu_{0}\omega\vec{H}, & \vec{\nabla}\times\vec{H} &= \sigma \vec{E} -i\epsilon_{0}\omega\vec{E},
\end{align}
where $\mu_{0}$ and $\epsilon_{0}$ are respectively the permeability and permittivity of vacuum. The charge density $\rho$ has been set equal to zero as we assume there are no net electrical charges in our setup.

The magnetic and electric field can be expressed as function of the vector potential $\vec{A}$. In the Weyl gauge, also called the temporal gauge, the electric potential $V$ vanishes per definition. As we can choose any gauge to describe the observable physics emanating from Maxwell's equations, we specifically opt for the Weyl gauge as this brings us as close as possible to the magnetostatics case. This is most appropriate when dealing with 1D quasi-stationary magnetic problems, as the one we are facing now.

Therefore one can write:
\begin{equation}
  \label{eq:11}
  \vec{H} = \frac{1}{\mu_{0}}\vec{\nabla}\times\vec{A}, \qquad \vec{E} = -i\omega\vec{A}.
\end{equation}
Substituting these equations in the Maxwell-Amp\`{e}re equation and using Gauss' law $\left(\vec{\nabla}\cdot \vec{A}= 0\right)$ we get:
\begin{equation}
  \label{eq:49}
  (\Delta - k_{i}^{2})\vec{A}_{i} = \vec{0}\qquad  k_{i}^{2}= -\omega^{2}\epsilon_{0}\mu_{0} + i\omega\mu_{0}\sigma_{i}.
\end{equation}
The real and imaginary parts of the parameter $k_{i}^{2}$ are respectively due to the displacement currents and the free currents. For low frequencies the real part is negligible with respect to the imaginary part, we therefore omit the displacement currents and $k_{i}^{2}$ becomes a purely imaginary number ($k_{i}^{2}=i\omega\mu_{0}\sigma_{i}$). This approximation is valid whenever $\omega\epsilon_{0}\ll\sigma_i$.

Exploiting the cylindrical symmetry, using separation of variables (with separation constant $\lambda$) and omitting the non-physical (exploding) solutions; the magnetic vector potential at coordinates $s, z$ can be written as follows:
\begin{subequations}
  \begin{align}
    \vec{A}_{0} &= \vec{e}_{\phi}\frac{m\mu_{0}}{4\pi}\int\limits_{0}^{\infty}f(\lambda)\exp{(-\lambda z)}J_{1}(\lambda s)\d \lambda + \frac{\mu_{0}}{4\pi}\frac{\vec{m}\times\vec{r}}{r^{3}},\label{eq:43}\\
    \vec{A}_{i} &= \vec{e}_{\phi} \frac{m\mu_{0}}{4\pi}\int\limits_{0}^{\infty}g_{i}(\lambda)\exp{(\gamma_{i} z)}\left[1 + x_{i}(\lambda)\exp{(-2\gamma_{i} z)}\right]J_{1}(\lambda s)\d \lambda,\\
    \vec{A}_{N} &= \vec{e}_{\phi} \frac{m\mu_{0}}{4\pi}\int\limits_{0}^{\infty}g_{N}(\lambda)\exp{(\gamma_{N} z)}J_{1}(\lambda s)\d \lambda.\label{eq:50}
  \end{align}
\end{subequations}
For ease of notation, we introduced the functions
\begin{equation}
  \gamma_{i}=\sqrt{\lambda^{2} + k_{i}^{2}}.
\end{equation}
The second part of $\vec{A}_{0}$ is the magnetic vector potential of an (ideal) magnetic dipole with moment $\vec{m}$. The functions $f(\lambda)$, $g_{i}(\lambda)$ and $x_{i}(\lambda)$ are dependent on the boundary conditions.

Applying the boundary condition $\vec{\nabla}\times\vec{A}=\vec{0}$ between the layers~\footnote{This ensures the absence of a discontinuity in the magnetic field, as required by the generally valid boundary conditions that follow from Maxwell's equations~\citep{jackson_classical_1975}. Indeed, since we do not expect highly conductive (metallic) layers in the upper earth, there are no boundary surface currents, the only possible source of discontinuities in $\vec H$, since we already set all magnetic permeabilities equal.}, we derive a recursion relation for $x_{i}(\lambda)$. Matching the air layer with the first soil layer using the same boundary condition results in  the function $f(\lambda)$:
\begin{align}
  f(\lambda) ={}& \lambda\frac{\gamma_{0} - Y_{1}}{\gamma_{0} + Y_{1}}\exp(-2\lambda h_{0}),\label{eq:12}\\
\intertext{where $Y_{1}$ is determined using the recursion relation:}
  Y_{i} \coloneqq{}& \gamma_{i} \frac{1-x_{i}\exp{(-2\gamma_{i}h_{i-1})}}
{1+x_{i}\exp{(-2\gamma_{i}h_{i-1})}}\\
        ={}& \gamma_{i} \frac{Y_{i+1} + \gamma_{i}\tanh(\gamma_{i}\Delta h_{i})} {\gamma_{i} + Y_{i+1}\tanh(\gamma_{i}\Delta h_{i})}.\label{eq:13}
\end{align}
The starting point of the recursion relation is determined from Equation~\eqref{eq:50}. Indeed, $x_{N}$ must be zero to obtain a physical magnetic field in the corresponding layer.
\section{Independent sheets}
\label{sec:independent-sheets}
\subsection{The LIN approximation}
\begin{figure}[t]
  \centering
  \begin{tikzpicture}[>=stealth]
    \def\m{.5}
    \def\w{3}
    \def\h{1.5}
    \def\dh{.5}

    \draw[->] (0,0) -- (0,\m) node [midway,right] {$\vec{m}$};

    \draw (-\w,-\h) -- (\w,-\h);
    \draw (-\w,-\h-\dh) -- (\w,-\h-\dh);

    \draw[<->] (\w,-\h) -- (\w, -\h-\dh) node [midway,right] {$\d h$};
    \draw[<->] (-\w,0) -- (-\w, -\h-\dh) node [midway,left] {$h$};

    \node at (0,{-\h*.5}) {Air};
    \node at (0,{-\h - \dh*.5}) {Thin sheet};
    \node at (0,{-\h - 2*\dh}) {Air};

  \end{tikzpicture}
  \caption{Independent sheet model. It consists of a conducting sheet floating in air. After integration with respect to the value $h$, one gets an approximation to the $N$-layer model.}\label{fig:mcneill}
\end{figure}

The LIN approach \citep{mcneill_electromagnetic_1980} considers a thin sheet at depth $h$ from the magnetic dipole with a conductivity $\sigma(h)$ and an infinitesimal thickness $\d h$ floating in air (see Figure~\ref{fig:mcneill}). Translating this to the setup of the previous section, we limit ourselves to a two-layer problem: the upper and lower half-space, both having a vanishing conductivity, and a thin layer in between. Denoting $\gamma_{1}$ as $\gamma$ we obtain:
\begin{align}
  Y_{1} &= \gamma\frac{\lambda+\gamma\tanh(\gamma\d h)}{\gamma + \lambda\tanh(\gamma\d h)} \\
        &\approx \lambda + k^{2}\d h,\\
  f(\lambda) &= -\frac{k^{2}\d h}{2}\exp{(-2\lambda h)}.
\end{align}

After calculating the integral in Equation~\eqref{eq:43} and taking the curl evaluated at $z$ equal to zero, we acquire the secondary fields a receiver on the same height as the dipole measures at a distance $s$~\citep{gradshteyn_table_1973}:
\begin{align}
  \label{eq:51}
  A_{\d h,\phi}(\vec{r}) &= -\frac{m\mu_{0}}{4\pi}\frac{k^{2}\d h}{2}\frac{\sqrt{s^{2} + {(2h+z)}^{2}} - 2h - z}{s\sqrt{s^{2} + {(2h + z)}^{2}}},\\
  H_{\d h, \hcp}(s\vec{e}_{s}) &= \frac{-m}{4\pi}k^{2}\d h\frac{h}{{\left(s^{2}+4h^{2}\right)}^{\nicefrac{3}{2}}},\label{eq:54}\\
  H_{\d h, \prp}(s\vec{e}_{s}) &= \frac{-m}{4\pi}\frac{k^{2}\d h}{2}\frac{s}{{(s^{2} + 4h^{2})}^{\nicefrac{3}{2}}}.\label{eq:55}
\end{align}
The horizontal coplanar system (HCP) corresponds with the vertical component of the magnetic field, while the perpendicular (PRP) system is the horizontal component due to a vertical dipole.

The actual problem we want to solve consists of a half-space with varying conductivity. Slicing the half-space in an infinite amount of thin sheets on top of each other, the secondary field can be obtained by integrating Equations~\eqref{eq:54} and~\eqref{eq:55} from zero to infinity with respect to the depth $h$. Using the dipole field $H_{D}$ at the same point as a normalisation coefficient, we define the normalised secondary field in terms of the so-called apparent conductivities $\sigma_{a,\hcp}$ and $\sigma_{a,\prp}$ \citep{mcneill_electromagnetic_1980}:
\begin{subequations}
  \label{eq:3}
  \begin{alignat}{4}
    \label{eq:53}
    h_{s,\hcp} \coloneqq{}& \frac{H_{s,\hcp}}{H_{D}} &&= \frac{i\omega\mu_{0}s^{2}}{4}\sigma_{a,\hcp}, &&\hspace{2cm} && \sigma_{a,\hcp} = \int\limits_{0}^{\infty}\sigma(\eta s)\frac{4\eta}{{(4\eta^{2} + 1)}^{\nicefrac{3}{2}}}\d \eta, \\
    h_{s,\prp} \coloneqq{}& \frac{H_{s,\prp}}{H_{D}} &&= \frac{i\omega\mu_{0}s^{2}}{4}\sigma_{a,\prp}, &&\qquad && \sigma_{a,\prp} = \int\limits_{0}^{\infty}\sigma(\eta s)\frac{2}{{(4\eta^{2} + 1)}^{\nicefrac{3}{2}}}\d \eta.
  \end{alignat}
\end{subequations}
In these equations we defined the dimensionless variable $\eta$ which is the depth of a layer $h$ normalised relative to the intercoil distance $s$. The first equation is the same as in McNeill, while the second one has been derived by \cite{saey_comparing_2015}.

\subsection{Shortcomings of the LIN approximation}
This approximation allows us to explain why we require LIN.\@ By summing the contributions of every thin sheet, we effectively eliminate the interactions between the thin layers. These interactions reduce the contribution of every layer and increase with the conductivity and, as such, the LIN model must break down. Moreover, a large intercoil distance increases the relative importance of the lower sheets. Their generated magnetic field contributions must hence travel a longer path through conductive matter, thereby decreasing their amplitude. The LIN approximation neglects this exponential dampening.

An expression for the low induction assumption can be obtained from the skin depth $\delta=\sqrt{\frac{2}{\omega\mu_{0}\sigma}}$. This material characteristic expresses how far electromagnetic fields can penetrate a material before its amplitude is considerably reduced. If the skin depth is much larger than the path the electromagnetic field has to traverse, one can remove the damping and the LIN approximation is valid. This path length is of the same order as the intercoil spacing $s$. Therefore, the LIN approximation is valid if $\frac{s}{\delta}\ll 1$, which can be rewritten as $\frac{\omega\mu_{0}\sigma s^{2}}{2}\ll 1$. This relation expresses the LIN assumption mentioned in the introduction.

\section{Introducing a conducting background: avoiding LIN}
\label{sec:intr-cond-backgr}

\begin{figure}[t]
  \centering
  \subfloat[A conductive sheet.]{%
    \begin{tikzpicture}[>=stealth,scale=.9]
      \def\m{.5}
      \def\w{3}
      \def\a{1.5}
      \def\h{1}
      \def\dh{.5}

      \draw[->] (0,0) -- (0,\m) node [midway,right] {$\vec{m}$};

      \draw (-\w,-\a) -- (\w,-\a);
      \draw (-\w,-\a-\h) -- (\w,-\a-\h);
      \draw (-\w,-\a-\h-\dh) -- (\w,-\a-\h-\dh);

      \draw[<->] (\w,0) -- (\w, -\a) node [midway,right] {$h_{0}$};
      \draw[<->] (\w,-\a-\h) -- (\w, -\a-\h-\dh) node [midway,right] {$\d h$};
      \draw[<->] (-\w,-\a) -- (-\w, -\a-\h-\dh) node [midway,left] {$h$};

      \node at (0,{-\a*.5}) {Air};
      \node at (0,{-\a-\h*.5}) {Conducting background};
      \node at (0,{-\a-\h - \dh*.5}) {Thin sheet $\sigma(h)$};
      \node at (0,{-\a-\h - 2*\dh}) {Conducting background};

    \end{tikzpicture}\label{fig:interactive_con}}
  \hspace{.1\textwidth}
  \subfloat[A thin sheet with no conductivity.]{%
    \begin{tikzpicture}[>=stealth,scale=.9]
      \def\m{.5}
      \def\w{3}
      \def\a{1.5}
      \def\h{1}
      \def\dh{.5}

      \draw[->] (0,0) -- (0,\m) node [midway,right] {$\vec{m}$};

      \draw (-\w,-\a) -- (\w,-\a);
      \draw (-\w,-\a-\h) -- (\w,-\a-\h);
      \draw (-\w,-\a-\h-\dh) -- (\w,-\a-\h-\dh);

      \draw[<->] (\w,0) -- (\w, -\a) node [midway,right] {$h_{0}$};
      \draw[<->] (\w,-\a-\h) -- (\w, -\a-\h-\dh) node [midway,right] {$\d h$};
      \draw[<->] (-\w,-\a) -- (-\w, -\a-\h-\dh) node [midway,left] {$h$};

      \node at (0,{-\a*.5}) {Air};
      \node at (0,{-\a-\h*.5}) {Conducting background};
      \node at (0,{-\a-\h - \dh*.5}) {Air};
      \node at (0,{-\a-\h - 2*\dh}) {Conducting background};
    \end{tikzpicture}\label{fig:interactive_non}}
  \caption{The interaction model. It consists of a dipole at a height $h_{0}$ above the ground. The ground is simulated as a thin sheet embedded in a conducting background. We subtract the contribution of a non-conductive sheet with the same dimensions. This eliminates the effect of the background. After integration w.r.t.\ the variable $h$, one gets an approximation of the $N$-layer model.}\label{fig:interactive}
\end{figure}

Introducing an interaction between the sheets allows us to reduce the LIN requirements, which will automatically lead to an improvement w.r.t.~the LIN model. We consider a sheet embedded in a half-space with fixed conductivity. Due to this half-space we introduce an interaction and thus dampening, while retaining the linear features of the problem. The system can be described as a three-layer model, with the upper and lower layer having the same conductivity $\sigma_{b}$. The middle layer has a conductivity $\sigma(h)$ and an infinitesimal thickness $\d h$ (see Figure~\ref{fig:interactive_con}). Using Equations~\eqref{eq:12} and~\eqref{eq:13} from Section~\ref{sec:solution-n-layer} and limiting ourselves to order one in $\d h$, we get:
\begin{align}
  \label{eq:56}
  Y_{3} &= \gamma_{b},\qquad Y_{2} \approx \gamma_{b} + (\gamma_{h}^{2}-\gamma_{b}^{2})\d h,\qquad Y_{1} \approx \gamma_{b} + (\gamma_{h}^{2}-\gamma_{b}^{2})\exp{(-2\gamma_{b} h)}\d h,\\
  f(\lambda) &\approx \lambda \frac{\lambda-\gamma_{b}}{\lambda+\gamma_{b}}\left[ 1 + 2\lambda \frac{\sigma(h)-\sigma_{b}}{\sigma_{b}}\exp{(-2\gamma_{b} h)}\d h\right] \exp{(-2\lambda h_{0})}.\label{eq:57}
\end{align}

In these calculations we included the effect of an a priori random background above and below the considered infinitesimally thin sheet. As eventually, we must again integrate over a continuum of such sheets, we need to remove this artificial surrounding background. In order to do so, we calculate the secondary field caused by the same setup but with a non-conductive thin sheet of air (see Figure~\ref{fig:interactive_non}). This leads to the same result as Equation~\eqref{eq:57} but with $\sigma(h)$ replaced by zero. After subtraction we obtain:
\begin{equation}
  \label{eq:52}
  \tilde{f}(\lambda)\approx 2\lambda^{2} \frac{\lambda-\gamma_{b}}{\lambda+\gamma_{b}} \frac{\sigma(h)}{\sigma_{b}}\exp{(-2\gamma_{b} h - 2\lambda h_{0})}\d h.
\end{equation}

The same approach as in the previous section is employed to calculate the magnetic field, yielding
\begin{align}
    h_{\d h, \genfrac{}{}{0pt}{}{\hcp}{\prp}} =& \frac{2s^{3}}{k_{b}^{2}} \frac{\sigma(h)}{\sigma_{b}}\d h \int\limits_{0}^{\infty} \lambda^{3}{(\lambda - \gamma_{b})}^{2} \exp{(-2\gamma_{b}h-2\lambda h_{0})} J_{\genfrac{}{}{0pt}{}{0}{1}}(\lambda s)\d \lambda,\label{eq:66}
\end{align}
These integrals have no analytic solution but are both numerically solvable.

For a stratified soil, the soil type may vary abruptly causing a jump in electrical conductivity. Therefore, the conductivity profile of the soil can be approximated as a (series of) step function(s). For this class of functions, the integration of Equation~\eqref{eq:66} with respect to $h$ is trivial. The secondary magnetic field is then:
\begin{align}
  \label{eq:68}
  h_{\genfrac{}{}{0pt}{}{\hcp}{\prp}}&= \sum\limits_{i}^{N}h_{i,\genfrac{}{}{0pt}{}{\hcp}{\prp}},\\
  h_{i, \genfrac{}{}{0pt}{}{\hcp}{\prp}} &= \frac{-s^{3}}{k_{b}^{2}} \frac{\sigma_{i}}{\sigma_{b}} \int\limits_{0}^{\infty} \frac{\lambda^{3}}{\gamma_{b}}{(\lambda - \gamma_{b})}^{2} \exp{(-2\lambda h_{0})}\left[\exp{\left(-2\gamma_{b}h\right)}\right]_{h_{i}}^{h_{i+1}} J_{\genfrac{}{}{0pt}{}{0}{1}}(\lambda s)\d \lambda.
\end{align}
Using the method of a digital filter described by~\cite{anderson_computer_1979}, these integrals are efficiently computable.  We have also tested the outcome of the numerical integration against a {\sf COMSOL}\textsuperscript{\textregistered} finite element simulation for the various considered profiles in this paper, finding perfect agreement, albeit at the cost of a considerably longer computation time for the simulations. However, using a simple approximation a reliable analytic estimate exists, as explained in the following section.

\subsection{Simplification of the interaction model}
\label{sec:simpl-inter-model}

To our knowledge, no analytic solution of Equation~\eqref{eq:66} exists,  but a simplification results in a closed form solution. From construction we expect the background conductivity to have the same order as the conductivity of soil. Due to this small value we can, using a Taylor expansion, approximate
\begin{equation}\label{extraeq}
\gamma_{b}\approx \lambda(1+0.5k_{b}^{2}\lambda^{-2}).
\end{equation}
For small $\lambda$, both integranda vanish due to the factor $\lambda^{3}$, while the Bessel functions are also well-behaved for small argument, see e.g.~\cite{abramowitz_handbook_1972}
\begin{equation}
J_0(x)=1+\mathcal{O}(x)\,,\qquad J_1(x)=\frac{x}{2}+\mathcal{O}(x^2).
\end{equation}
As such, the major contribution to the integrals~\eqref{eq:66} will come from the $\lambda$-not-so-small-region, which underpins using the approximation \eqref{extraeq} under the integral sign. Notice that the potentially compensating large value of the intercoil distance $s$ does not spoil this picture, since both Bessel functions $J_{0,1}(\lambda s)$ essentially behave as $\frac{1}{\sqrt{\lambda s}}$ for $\lambda s\gg 1$, which does not eliminate the dominant $\lambda^3$-prefactor at small $\lambda$.

Thus, applying the prescribed Taylor approximation on the polynomial in the integrandum of Equation~\eqref{eq:66} and assuming a dipole lying on the ground ($h_{0}=0$) yields~\citep{gradshteyn_table_1973}:
\begin{align}
  h_{\d h,\hcp} \approx& \frac{s^{3}}{k_{b}^{2}} \frac{\sigma(h)}{\sigma_{b}} \int\limits_{0}^{\infty} \lambda^{3}{\left(\frac{k_{b}^{2}}{2\lambda}\right)}^{2} \exp{(-2\gamma_{b}h)} J_{0}(\lambda s)\d \lambda \nonumber \\
          =& \frac{i\omega\mu_{0}\sigma(h)s^{2} \d \eta}{4} 4\exp{(-k_{b}s\sqrt{4\eta^{2} + 1})}\frac{\eta}{4\eta^{2} + 1}\left(k_{b}s + \frac{1}{\sqrt{4\eta^{2} + 1}}\right),\\
    h_{\d h, \prp} \approx& \frac{i\omega\mu_{0}\sigma(h)\rho^{2} s\d h}{4} 2\frac{\partial^{2}T(s, 2h)}{\partial (2h)\partial s},\\
  T(s, z) =& I_{0}\left[\frac{ks}{2}(\sqrt{s^{2}+z^{2}}-z)\right]K_{0}\left[\frac{ks}{2}(\sqrt{s^{2} + z^{2}}+z)\right].\\
  \intertext{Assuming a step function as conductivity profile, the $i$\textsuperscript{th} layer causes the following magnetic field:}
\label{eq:71b}  h_{i,\hcp} \approx& -\frac{i\omega\mu_{0}\sigma_{i}s^{2}}{4}{\left[ \frac{\exp{-k_{b}s\sqrt{4\eta^{2}+1}}}{\sqrt{4\eta^{2}+1}}\right]}_{\eta_{i}}^{\eta_{i+1}},\\
  h_{i,\prp} \approx& \frac{i\omega\mu_{0}\sigma_{i}s^{2}}{4} {\left[\frac{k_{b}s}{2\sqrt{4\eta^{2}+1}}\left(I_{1}(r_{-})K_{0}(r_{+}) - I_{0}(r_{-})K_{1}(r_{+})\right)\right]}_{\eta_{i}}^{\eta_{i+1}},
\end{align}
where,
\begin{equation}
  \label{eq:71}
  \eta_{i} = \frac{h_{i}}{s},\qquad  r_{i,\pm} = \frac{k_{b}s}{2}\left(\sqrt{4\eta_{i}^{2} + 1} \pm 2\eta_{i}\right).
\end{equation}

This method combines the advantages of the two earlier developed models. Due to the closed form solution it has the simplicity of the LIN model. On the other hand, the dampening from the interaction model considerably reduces the LIN requirements. We refer to this approximation as the damped model.

\subsection{The optimal background conductivity}
\label{sec:optim-backgr-cond}
\begin{figure}[t!]
  \centering
  \includegraphics[width=.5\textwidth]{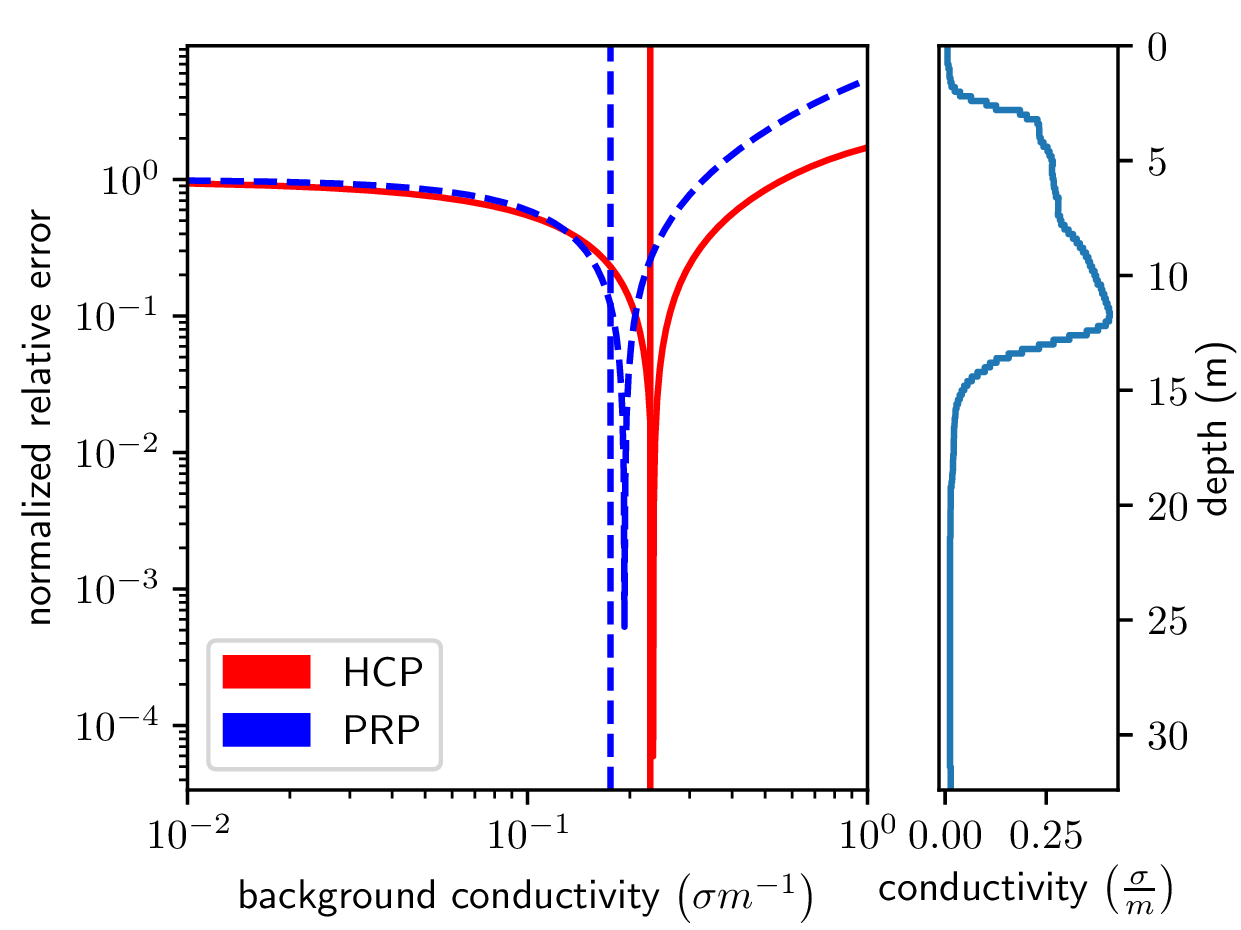}
  \caption{The relative error of the damped model as a function of the background conductivity. We normalised the error with the relative error of the LIN model. The vertical lines indicate the apparent conductivity. The frequency of the dipole is \SI{1.6}{\kilo\hertz} and the intercoil spacing is \SI{20}{\meter}. On the right, the vertical conductivity profile is plotted. The profile is based on the data of~\cite{hermans_imaging_2012}.}\label{fig:loopBack}
\end{figure}

So far, the damped model requires a dipole lying on the ground and an unknown parameter $\sigma_{b}$. The first restriction can be overcome using a shift in the variable of the function $\sigma(h)$. Indeed, if we define a new conductivity profile of the following form:
\begin{equation}
  \tilde{\sigma}(h) =
  \begin{cases}
    0 & 0<h<h_{0}\\
    \sigma(h-h_{0}) & h_{0}<h
  \end{cases},\label{eq:59}
\end{equation}
then the dipole rests a distance $h_{0}$ above the ground, with the top (air) layer, of thickness $h_{0}$, having a zero conductivity.

The parameter $\sigma_{b}$ can be set equal to a fixed value. In that case we have a completely linear forward model. Unfortunately, pinpointing the exact value is a difficult task. In Figure~\ref{fig:loopBack}, the normalised relative error (the norm is the relative error of the LIN solution) is plotted as function of the background conductivity. For a vanishing background conductivity the solution, as one would expect, converges to the LIN solution. A small deviation w.r.t.\ the optimal value results in a large difference in error. Furthermore, the apparent conductivity is not always a correct indication of the optimal background conductivity.

An alternative approach to determine the background conductivity is to use some of the knowledge we know (or acquire) about our system. The damping is substantially caused by the layers on top of the considered layer. Thus if we calculate the secondary magnetic field caused by the $i^{\text{th}}$ layer we can approximate $\sigma_{b}$ for this layer as the weighted average of the conductivities of the layers on top of this layer. As weights we choose the thicknesses of the corresponding layers. Using this method, every layer has a different background conductivity:
\begin{equation}
  \label{eq:2}
  \sigma_{b, i}= \frac{\sum_{j=1}^{i}\sigma_{j}\Delta h_{j}}{\sum_{j=1}^{j}\Delta h_{j}}.
\end{equation}

In case the thickness of the $i^{\text{th}}$ layer would be quite large, we neglect the damping in this layer, especially for the lower regions. To overcome this, we simply split thick layers into thinner sublayers. In later work, where we will discuss the inverse problem in which case any preknowledge of the number of layers or their respective conductivities is missing, we will have to assume a sufficiently large set of very thin layers. This type of procedure automatically allows us to model the conductivity profile $\sigma$ as a series of step functions.

\section{Comparison between the damped model and the exact iterative solution}
\label{sec:comp-two-meth}

To compare our new model we consider two systems: one with high and one with low conductivity. For two reasons, only the imaginary part of the secondary field is considered. First of all the LIN result has no real part, and therefore a direct comparison is not possible. Secondly, the magnetic field due to the source dipole is real and substantially larger than the secondary field in magnitude. Experimental (or even numerical) measurements with high precision are therefore quite difficult. For all systems we use a dipole with a frequency of \SI{1.6}{\kilo\hertz}, while for the intercoil distance we take up to \SI{20}{\meter}. These values are based on a typical configuration of the GEONICS EM34 system.

\begin{figure}[t!]
  \centering
  \subfloat[A conductivity sounding of a stratified resistive soil\label{fig:profLuik}]{%
   \includegraphics[width=.48\textwidth]{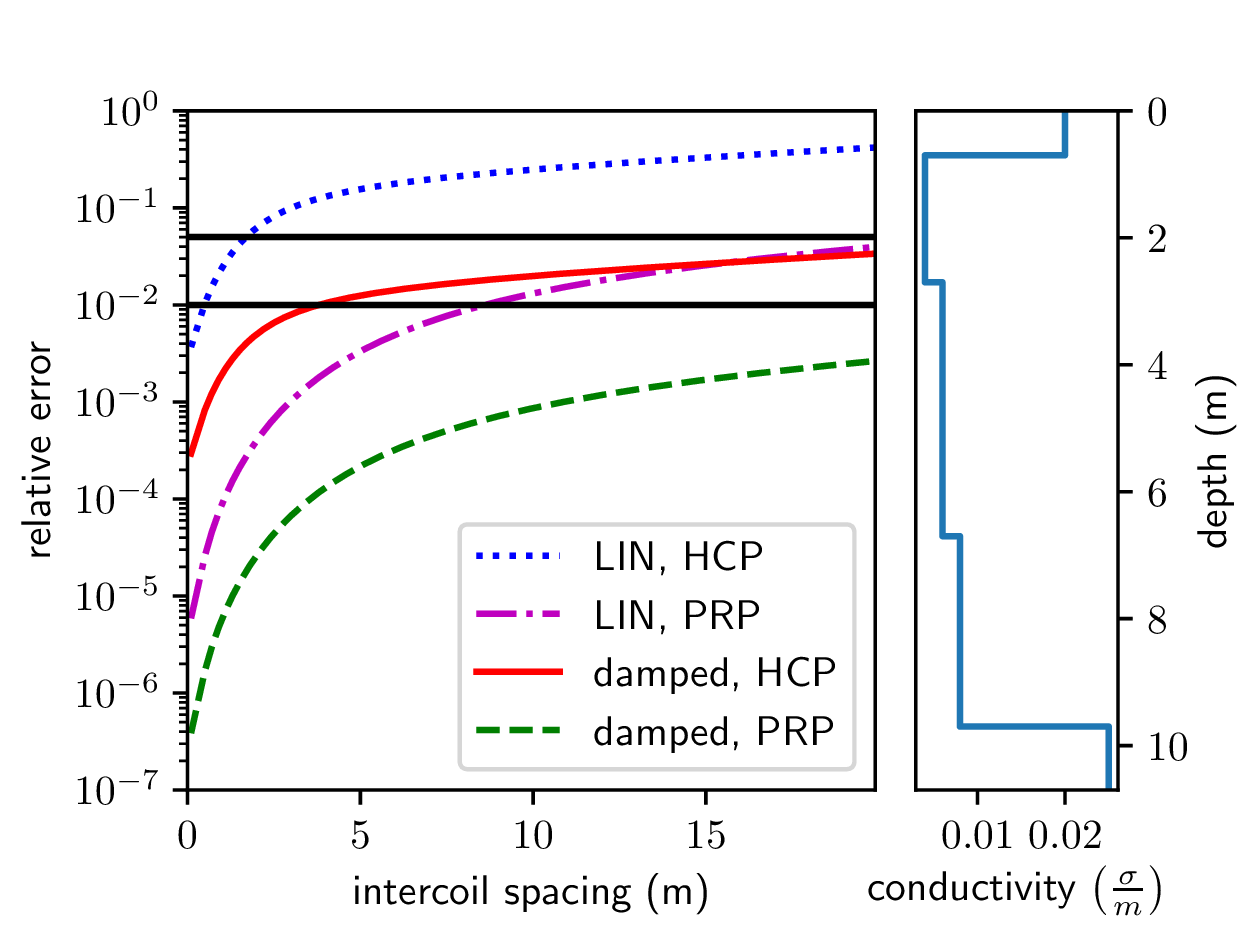}}
  \quad
  \subfloat[A conductivity sounding of a stratified conductive soil\label{fig:profP11}]{%
    \includegraphics[width=.48\textwidth]{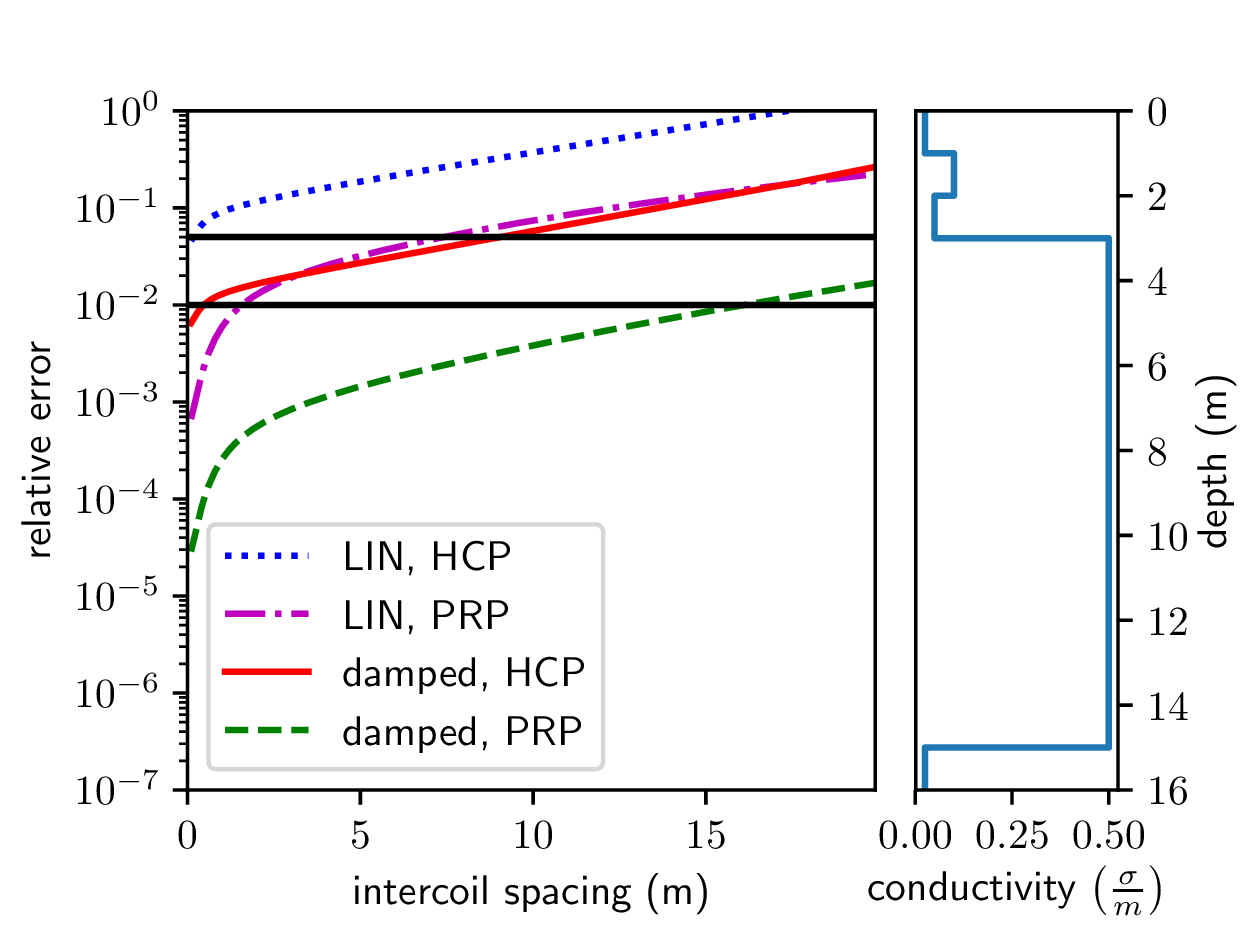}}\\
  \subfloat[A conductivity sounding based on borehole data from~\cite{hermans_facies_2017}]{\includegraphics[width=.48\textwidth]{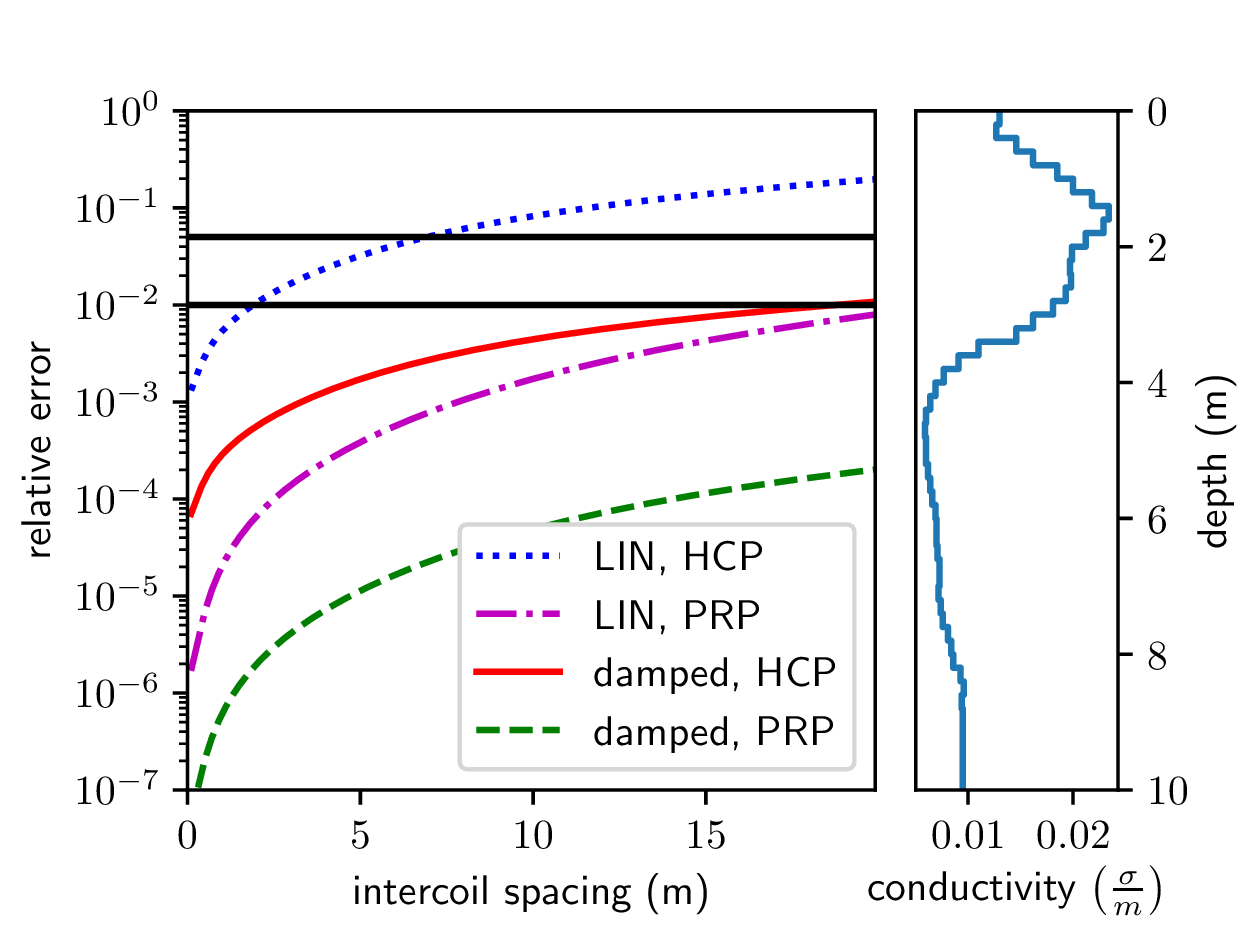}}
  \quad
  \subfloat[A conductivity sounding based on borehole data from~\cite{hermans_imaging_2012}. Due to the presence of saltwater the conductivity increases sharply.]{\includegraphics[width=.48\textwidth]{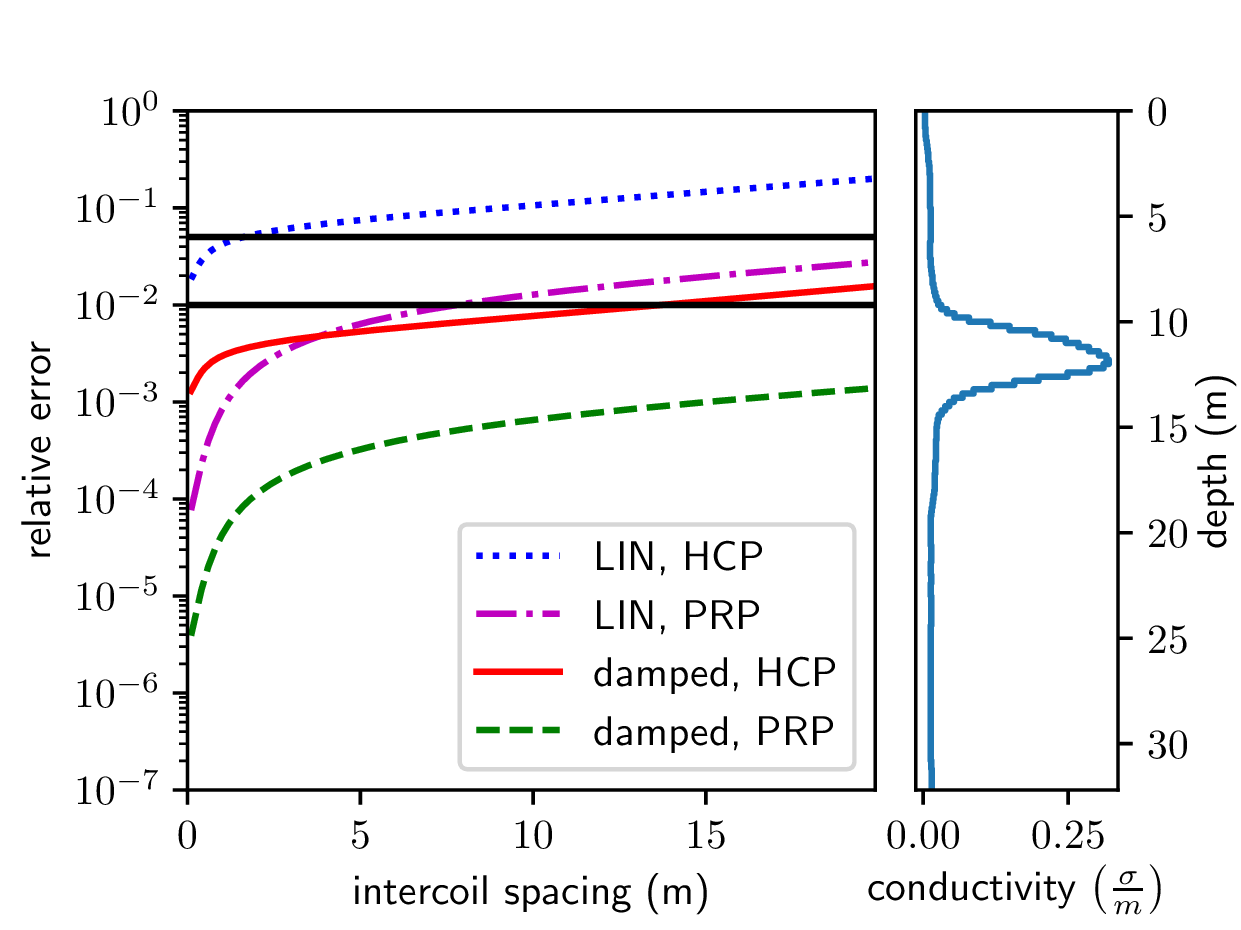}}
  \caption{The relative error on the imaginary part of the secondary magnetic field. The exact result is obtained from the theoretical result (see Equation~\eqref{eq:43} with $f(\lambda)$ determined from Equation~\eqref{eq:12}). On the right of each figure, the conductivity profile is plotted. The horizontal black lines correspond to the 1\% and 5\% error.\label{fig:loopRho}}
\end{figure}

\begin{figure}[t]
  \centering
  \subfloat[]{%
   \includegraphics[width=.48\textwidth]{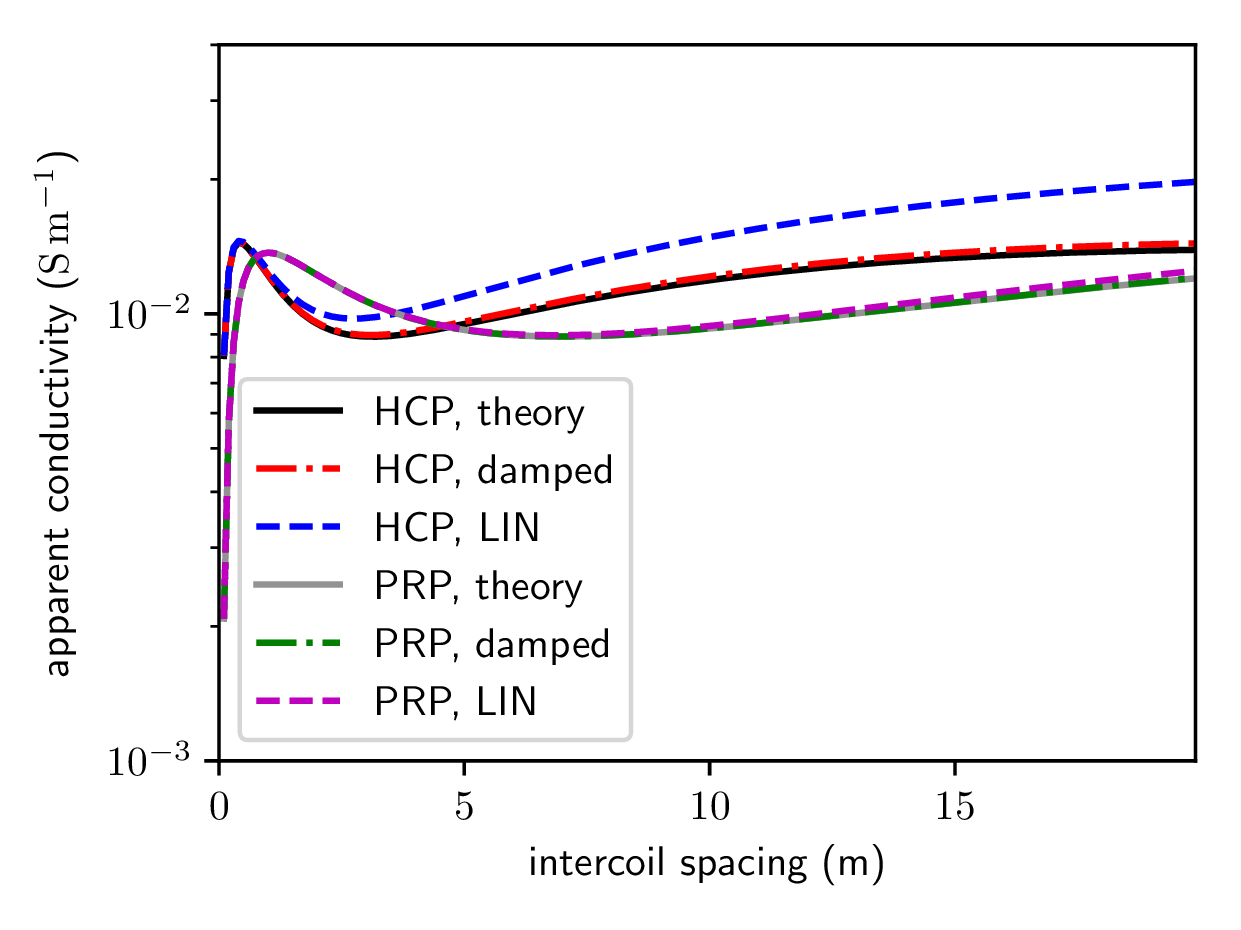}}
  \quad
  \subfloat[]{%
    \includegraphics[width=.48\textwidth]{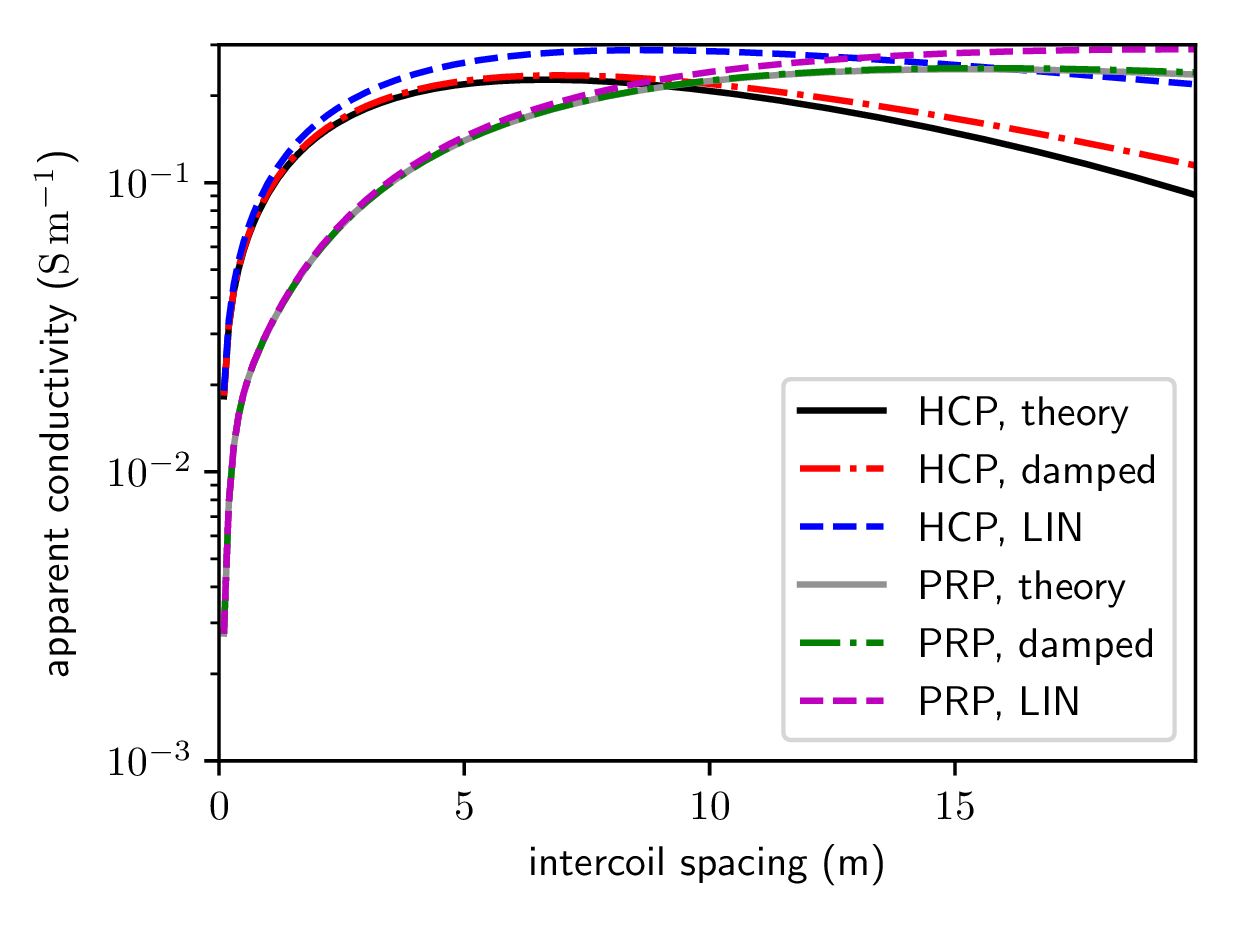}}\\
  \subfloat[]{\includegraphics[width=.48\textwidth]{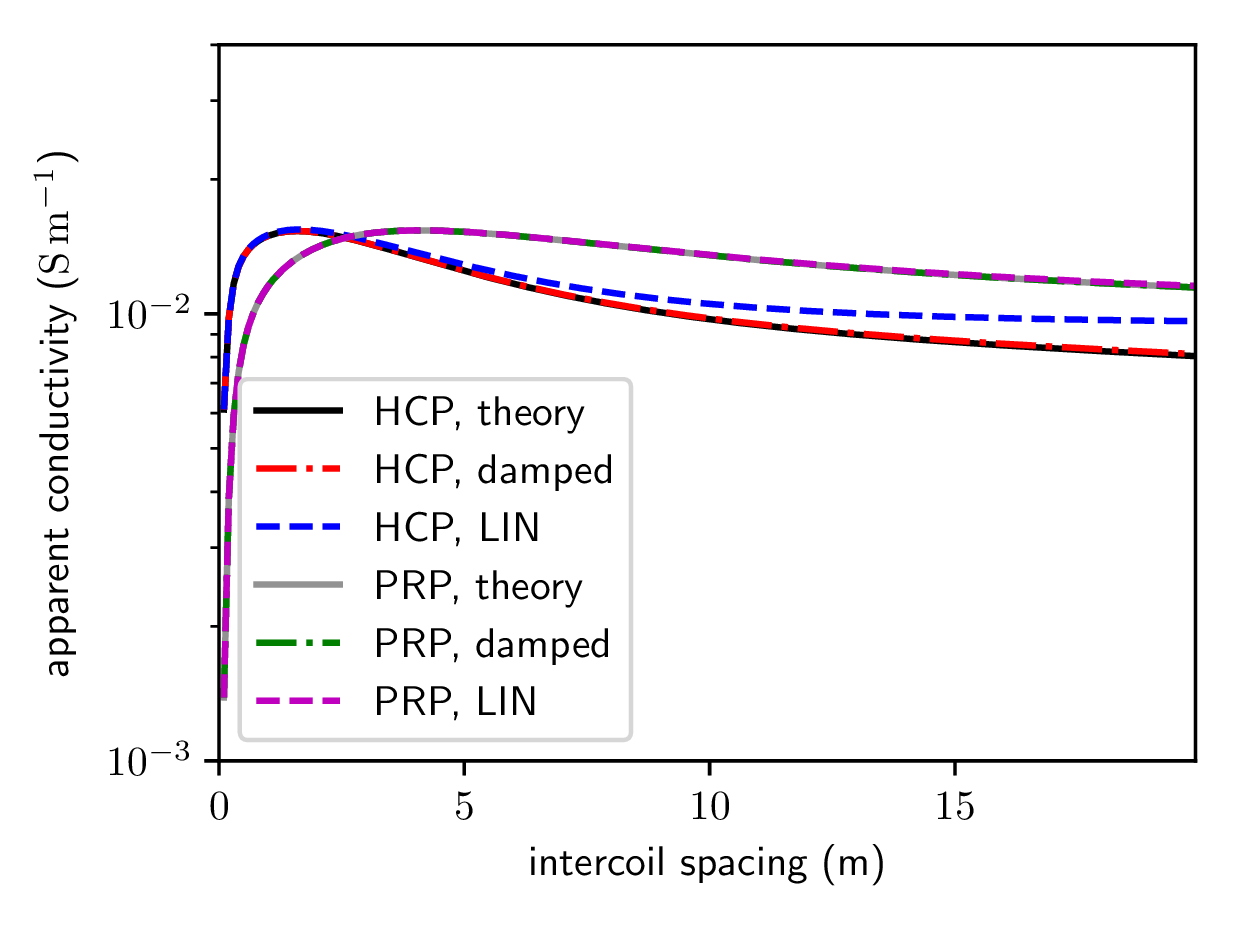}}
  \quad
  \subfloat[]{\includegraphics[width=.48\textwidth]{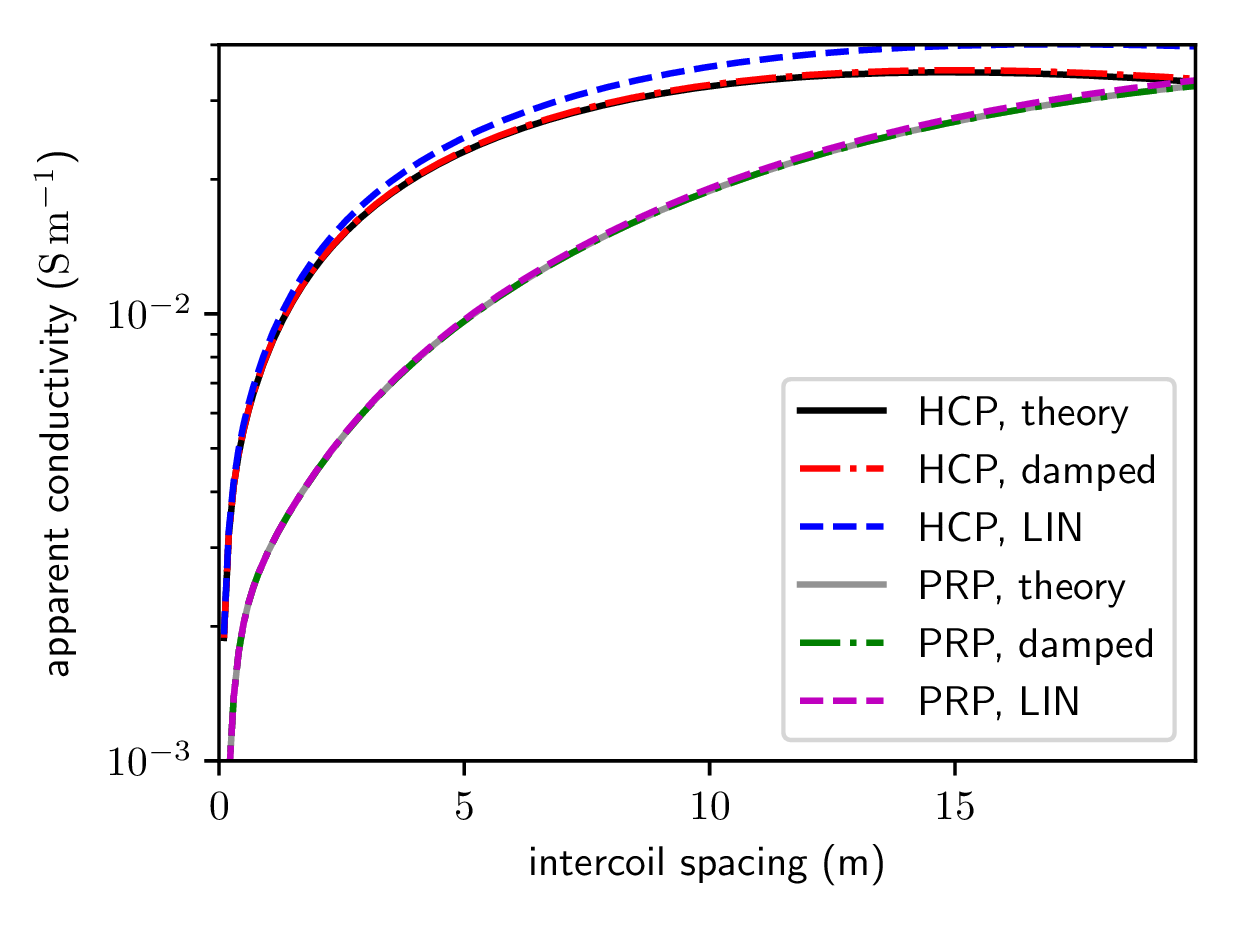}}
  \caption{The apparent conductivity for the four soundings from Figure~\ref{fig:loopRho}. The apparent conductivity was calculated based on Equation~\eqref{eq:3} with the relative magnetic field calculated from the three models (LIN, damped and theory). Therefore, this is the value which most measurement devices would return.\label{fig:loopRhoApparent}}
\end{figure}
\begin{figure}[t]
  \centering
  \subfloat[The LIN model]{%
    \includegraphics[width=.48\textwidth]{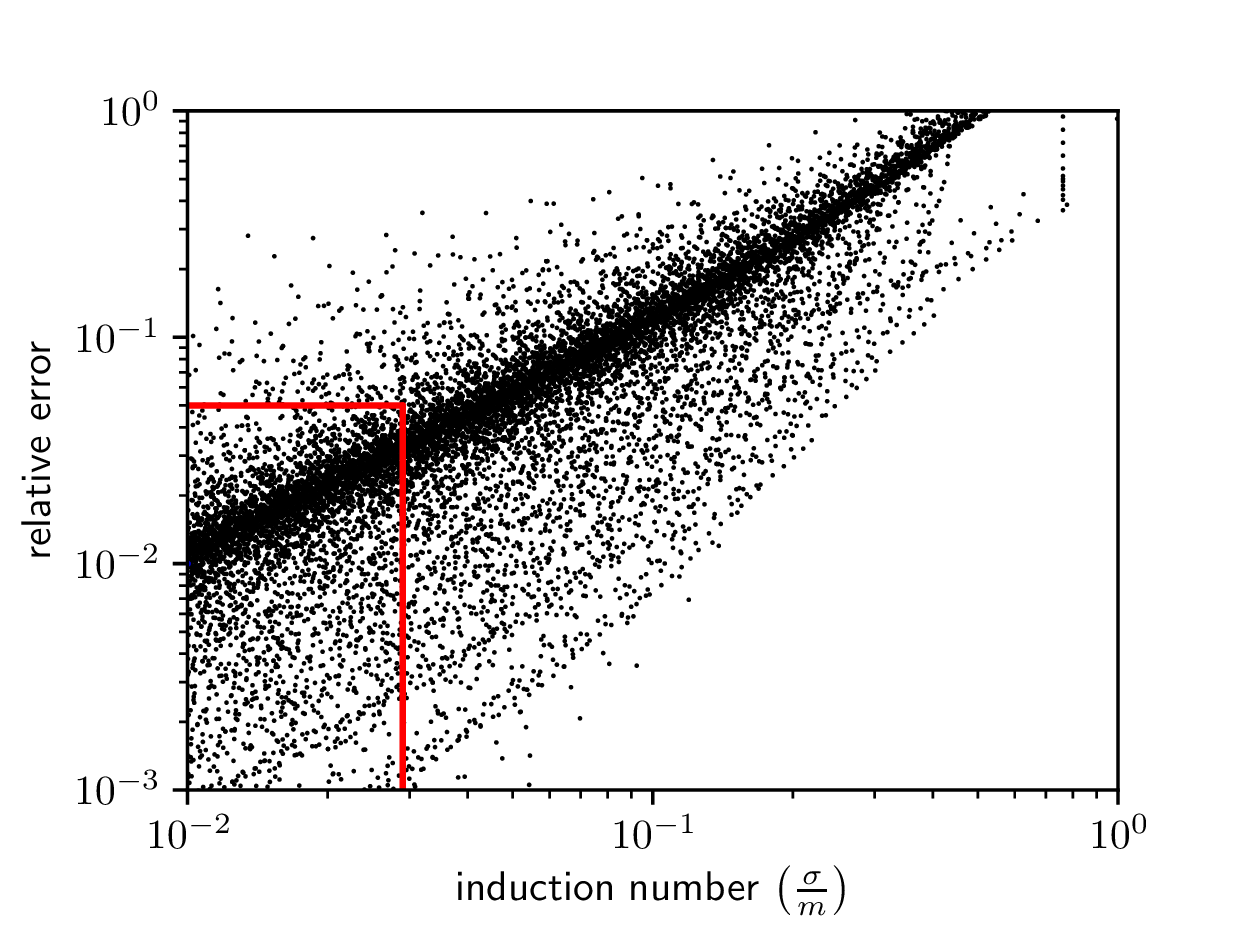}}\quad
  \subfloat[The damped model]{%
  \includegraphics[width=.48\textwidth]{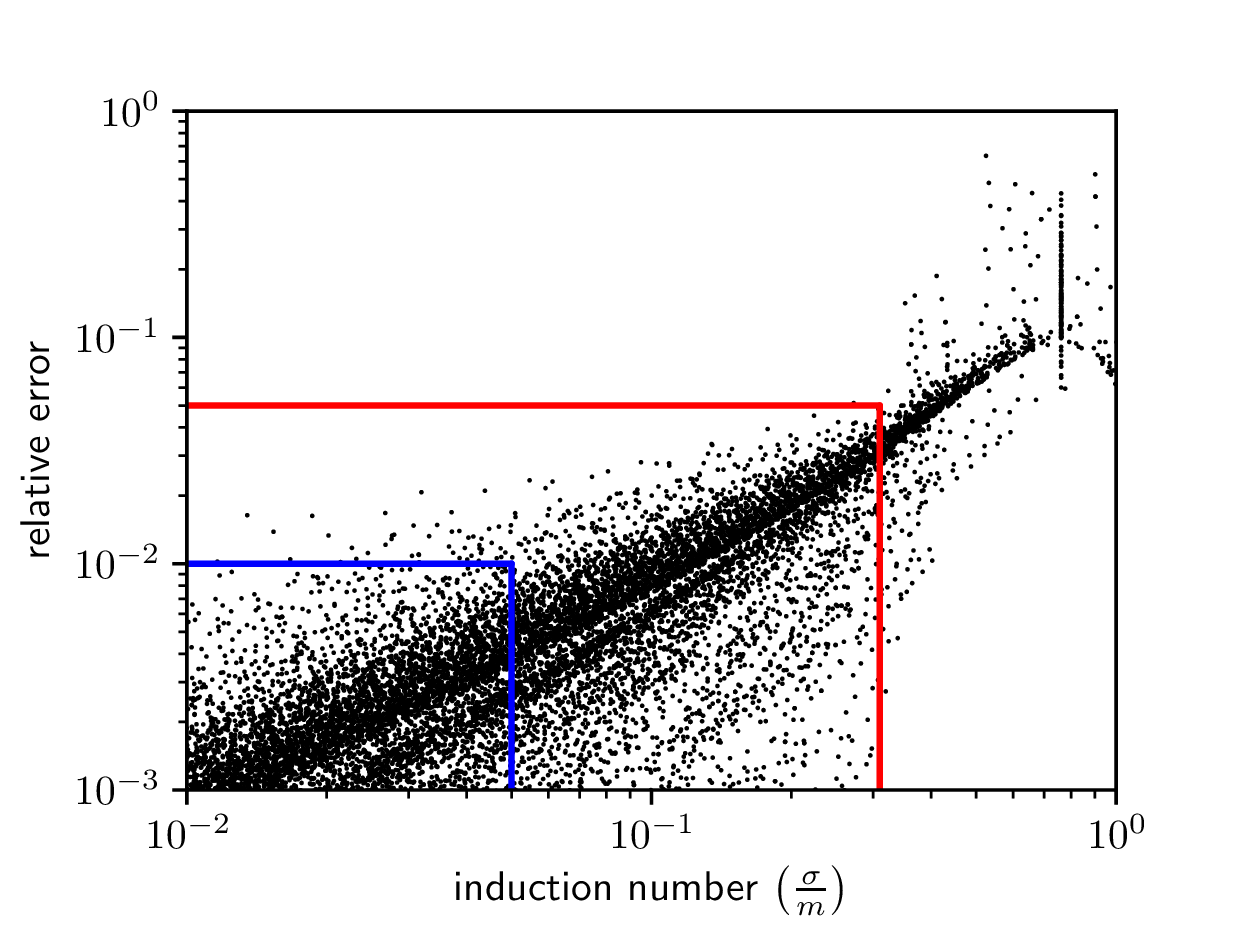}}
  \caption{The relative error of the LIN model (left) and the damped model w.r.t.\ the theoretical solution in function of the induction number. We considered two layered soils and soils from experimental measurements. The horizontal lines correspond to the 1\% and 5\% error.}\label{fig:loopInduction}
\end{figure}

In Figure~\ref{fig:loopRho} and~\ref{fig:loopRhoApparent} the relative error of the magnetic field and the apparent conductivity are plotted, respectively. We considered four profiles, two discontinuous and two continuous ones. Each type has two profiles representing a conductive and a resistive soil. For all four profiles we can report that our model has an error almost ten times smaller than the LIN error. Not only is the error smaller than the LIN model, it also remains smaller than 5\% for intercoil distance below 10~m. This percentage corresponds to the measurement error (see e.g.~\cite{gf_instruments_gf_2018} or~\cite{geonics_limited_geophysical_2018}). For more resistive soils the error remains smaller than 1\% for intercoil distances lower than 10~m. We also observe that the error is larger for the stratified soils than for a continuous profile.

As expected, the error increases with intercoil distance and conductivity. To test the limits of our model, the relative error is plotted function of the induction number $\sqrt{\frac{\omega\mu\sigma_{a}}{2}}s$ (see Figure~\ref{fig:loopInduction}). To obtain this plot we considered different soils (conductivity profiles based on borehole data and artificial two layered soils with $\sigma < \SI{0.5}{\siemens\per\meter}$), frequencies ($f < \SI{1.6}{\kilo\hertz}$) and intercoil distances ($s < \SI{20}{\meter}$); the three parameters defining the induction number. If we demand an error smaller than 1\%, the induction number should be smaller than 0.004 and 0.05 respectively for the LIN and the damped model. The 5\% error corresponds with an upper limit of 0.029 and 0.31 respectively for the LIN and damped model. Summarizing, the latter performs an order of magnitude better than the former. We notice that these maximal induction numbers are conservative estimates, as will become clear for example from the hydrogeological example in Section~\ref{sec:salt}.

\subsection*{Vertical sensitivity}

A good model not only predicts the secondary field accurately, it should also assign a correct weight to each layer. This, called the vertical sensitivity, can be used to determine to what depth the soil interacts with the emitter, which is essential information for the inversion of data from EMI surveys. To determine this function at a certain depth we calculate the secondary field for a system consisting of soil above that depth while beneath it is a non-conductive half-space (air). We calculate this curve for two intercoil distances (\SI{5}{\meter} and \SI{10}{\meter}) and for different conductivities. We normalise the results with the secondary field in case only a soil is present (i.e., no air below it).

\begin{figure}[t!]
  \centering
  \subfloat[HCP field\label{fig:doeLuikz}]{\includegraphics[width=.48\textwidth]{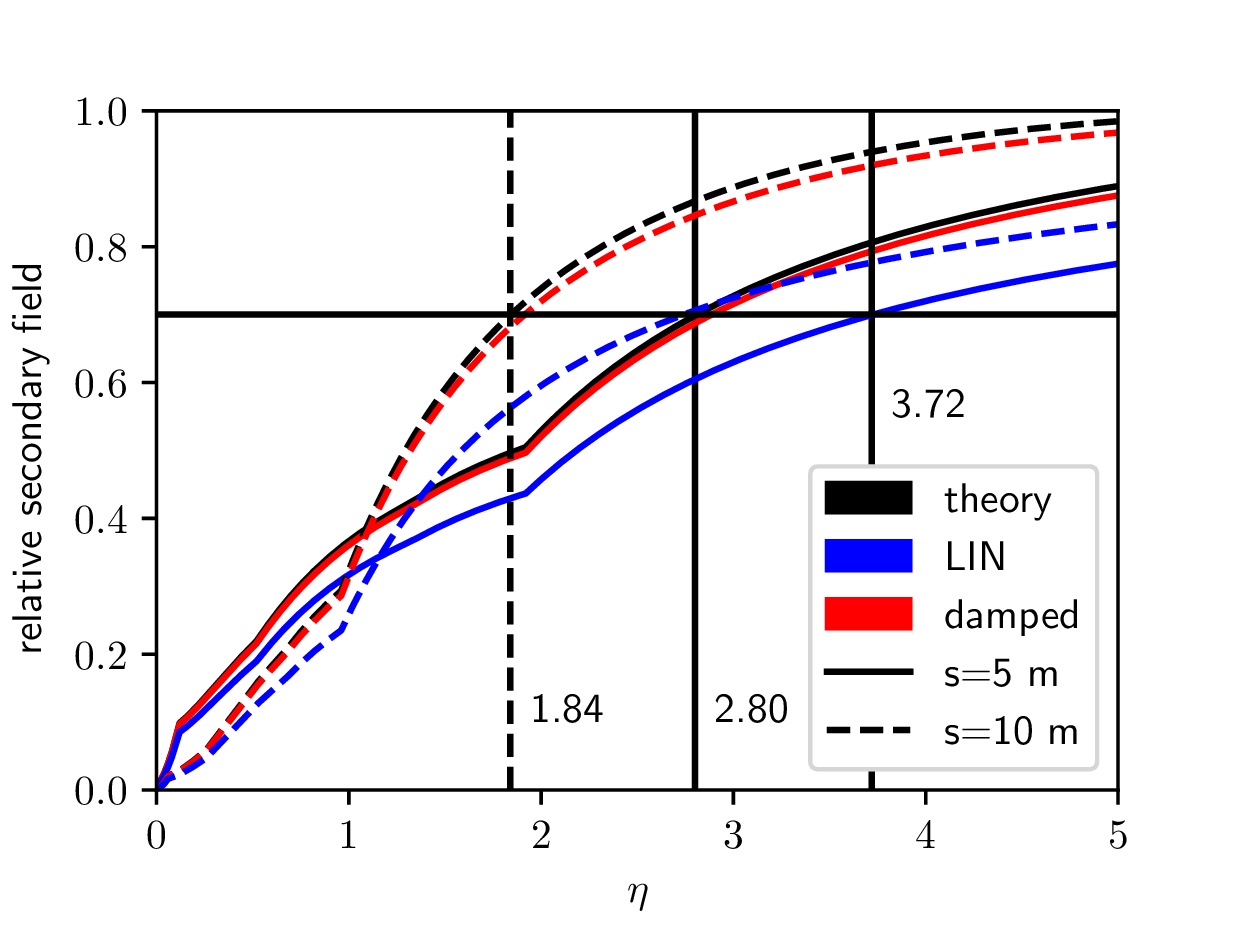}}\hfill
  \subfloat[HCP field\label{fig:doeP11z}]{%
    \includegraphics[width=.48\textwidth]{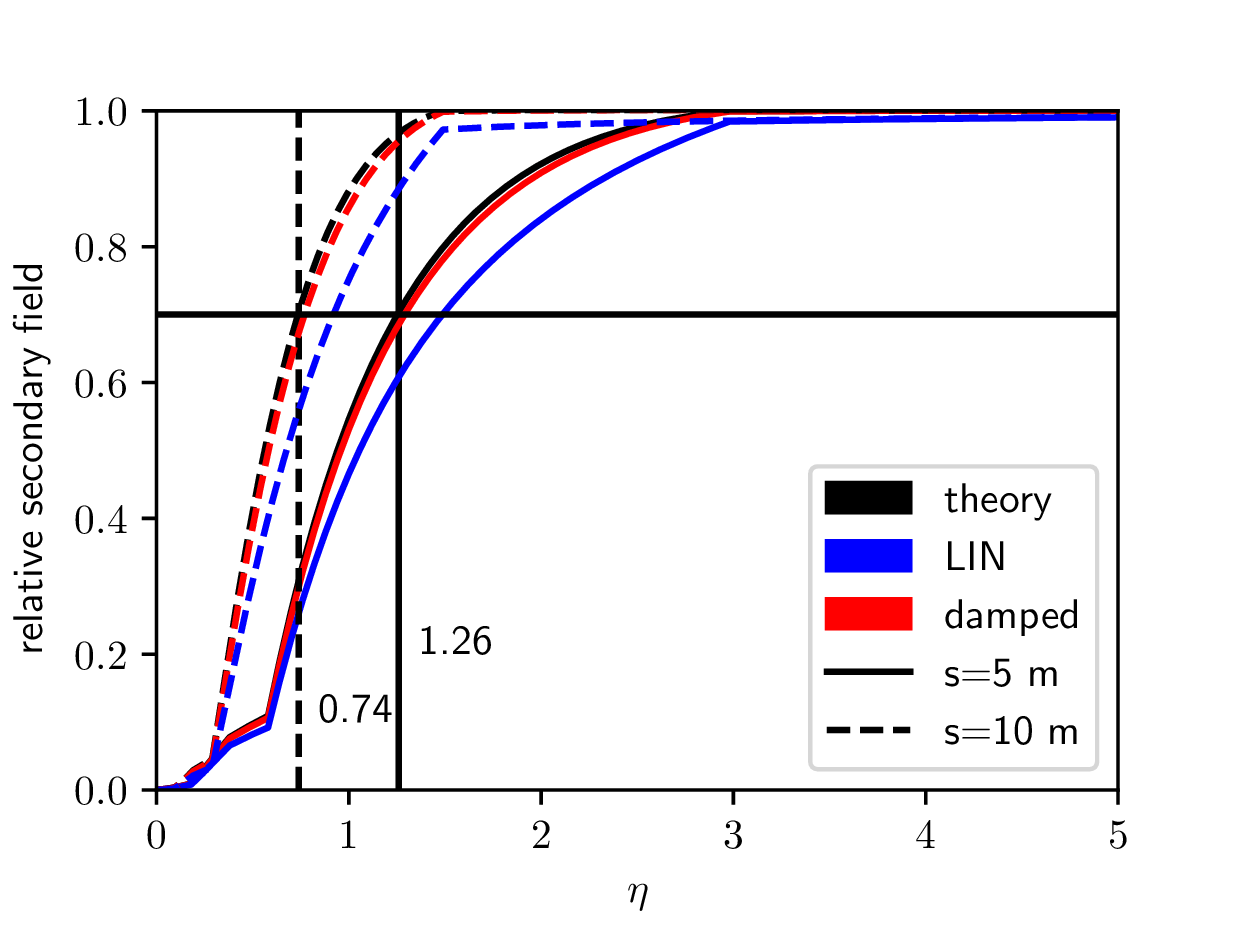}}\\
  \subfloat[PRP field\label{fig:doeLuikr}]{%
    \includegraphics[width=.48\textwidth]{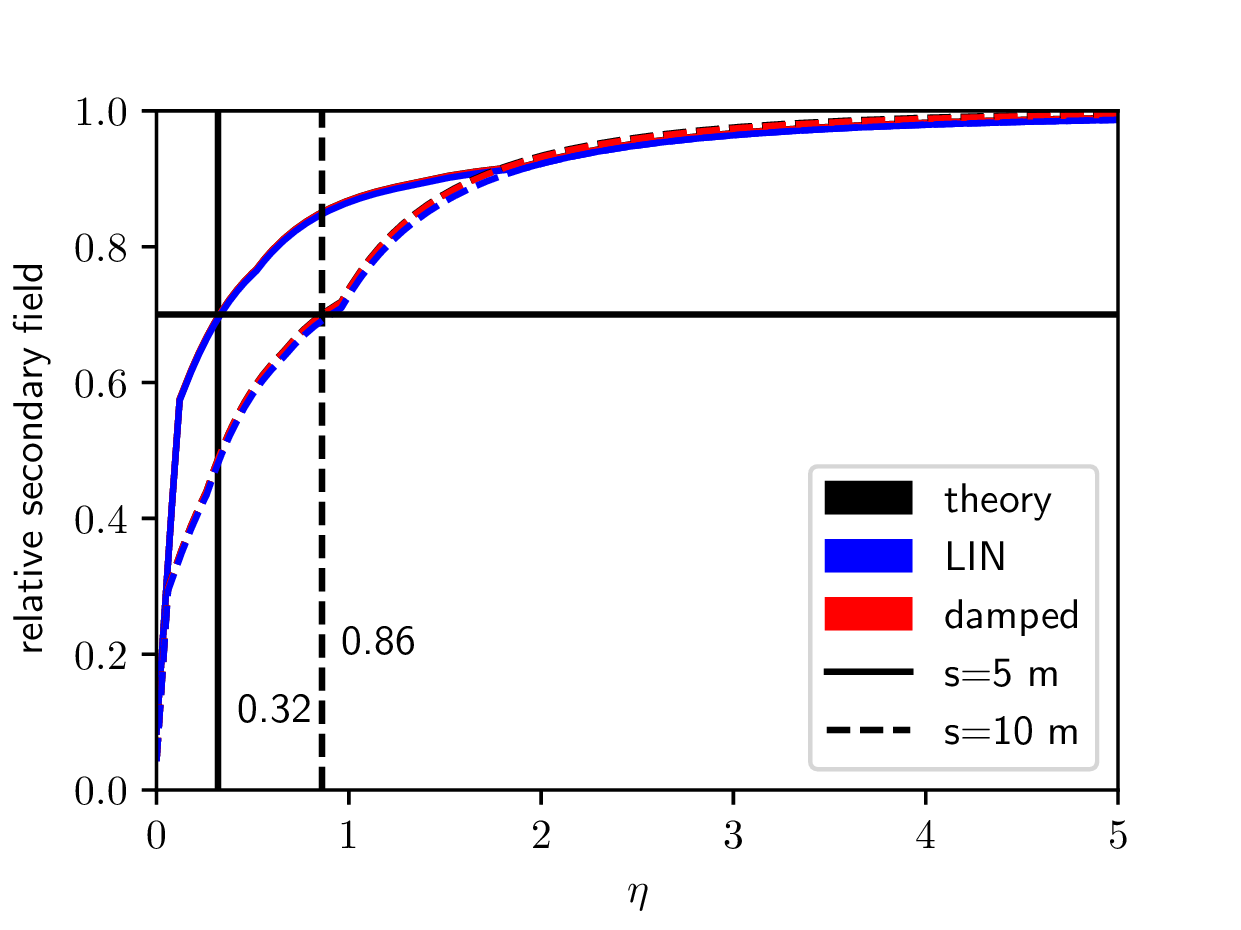}}\hfill
  \subfloat[PRP field\label{fig:doeP11r}]{%
    \includegraphics[width=.48\textwidth]{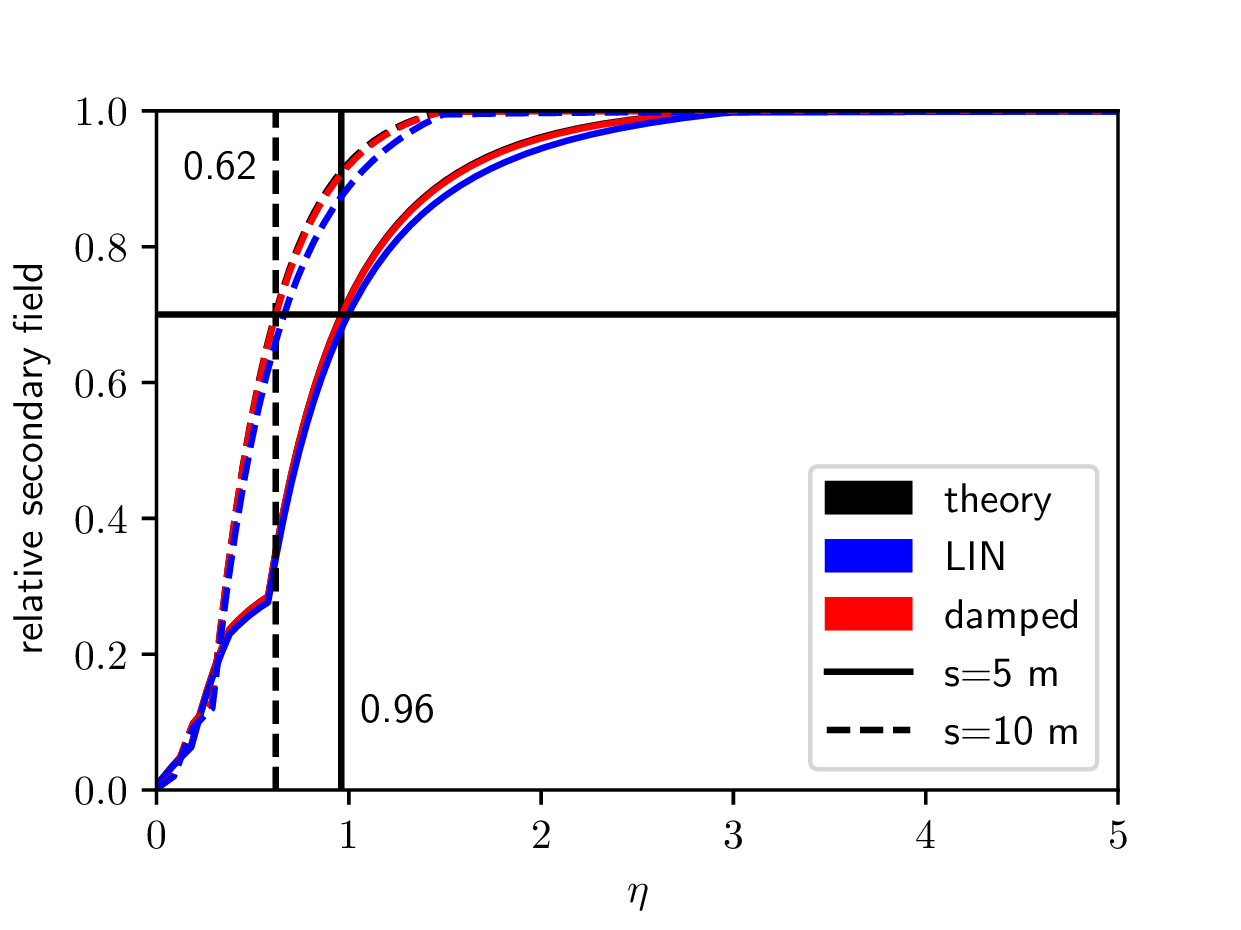}}
  \caption{Relative sensitivity for the normalised depth for the exact solution (black), the LIN model (blue line) and the damped model (red line). Sensitivity was calculated for intercoil distances $s= \SI{5}{\meter}$ (solid line) and $s= \SI{10}{\meter}$ (dashed line). For the PRP orientation, both models are a very good approximation of the theory (the black line is obscured by the red line). The sensitivity of the HCP orientation is only well approximated by the damped model. This can be explained from the fact that this field is mainly caused by the deeper parts of the soil. The conductivity profile of~\ref{fig:doeLuikz} and~\ref{fig:doeLuikr} is plotted in Figure~\ref{fig:profLuik}, the profile of~\ref{fig:doeP11z} and~\ref{fig:doeP11r} is plotted in Figure~\ref{fig:profP11}.}\label{fig:loopDoe}
\end{figure}

For conductivities around \SI{20}{\milli\siemens\per\meter} (see Figure~\ref{fig:doeLuikz} and~\ref{fig:doeLuikr}) the discrepancy between the LIN model and theory is already large for the HCP field. In case of the PRP field, the difference is minimal. For larger intercoil distances the error, as one would expect, increases. For both components and separations, our model follows the theory almost exactly. As the LIN model respectively underestimates and overestimates the upper and lower soil, we can expect it to perform badly for soils with a strongly varying conductivity.

The conclusion for the sensitivity is almost the same in case of saline soils (see Figure~\ref{fig:doeP11z} and~\ref{fig:doeP11r}). Our model describes the theory very well, while the LIN model deviates from the theoretical curve, especially for the HCP field and for larger intercoil distances. The only effect of a higher conductivity is the shift of the curves to the left. As our model has a sensitivity almost completely similar to the theory, we expect it to perform very well for the detection of the interface between layers.

The depth of exploration ($d_{e}$) was introduced by McNeill as the depth where 70\% of the secondary field is caused by the soil above $d_{e}$\footnote{This definition is not explicitly mentioned, but \cite{callegary_vertical_2007} deduced it from the given $d_{e}$'s.}. In Figure~\ref{fig:loopDoe} this is indicated with a horizontal black line, and the corresponding $\eta_{e}$ for the different models is also mentioned. As before, $\eta$ is the depth rescaled relative to the intercoil separation. A $d_{e}$ of $1.5 * s$ is mentioned in McNeill. The latter value is however only correct for a homogeneous half-space, in the non-homogeneous case we expect deviations. Using the same definition a $d_{e}$ of $0.49 * s$ can be calculated for the horizontal component. The obtained values are a good indication but as $d_{e}$ does not change with conductivity, they are not broadly valid. In our results a decrease in $d_{e}$ is found for increasing conductivities, especially for larger $s$. This is consistent with results from the literature: from surveys \cite{saey_electrical_2012} found a 25\% decrease in $d_{e}$, while \cite{callegary_vertical_2007}, using finite element simulations, calculated a decrease up to 50\%.

\section{A simple saltwater infiltration model}
\label{sec:salt}

\begin{figure}[h]
  \centering
  \subfloat[ ]{%
    \includegraphics[width=.48\textwidth]{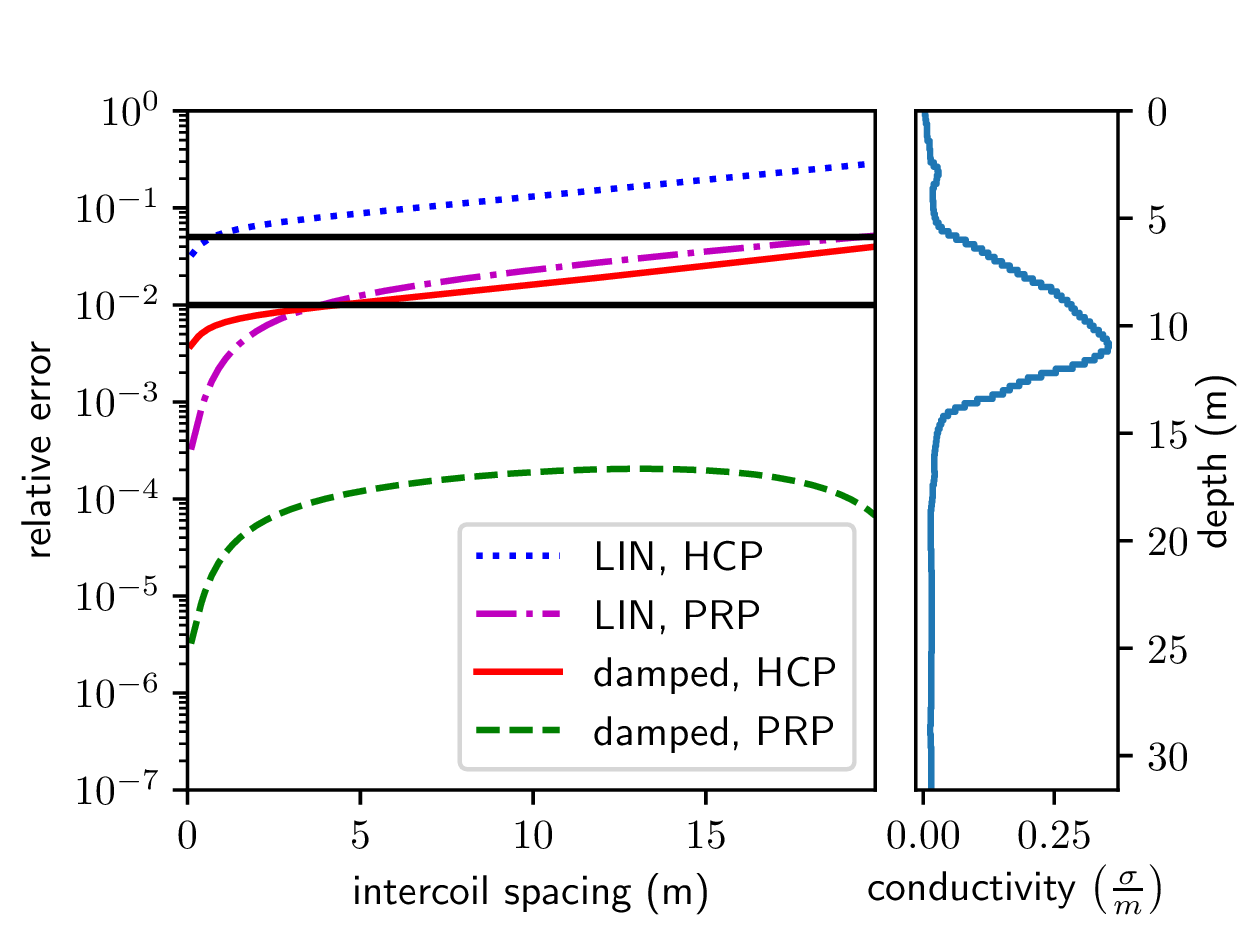}}\hfill%
  \subfloat[ ]{%
  \includegraphics[width=.48\textwidth]{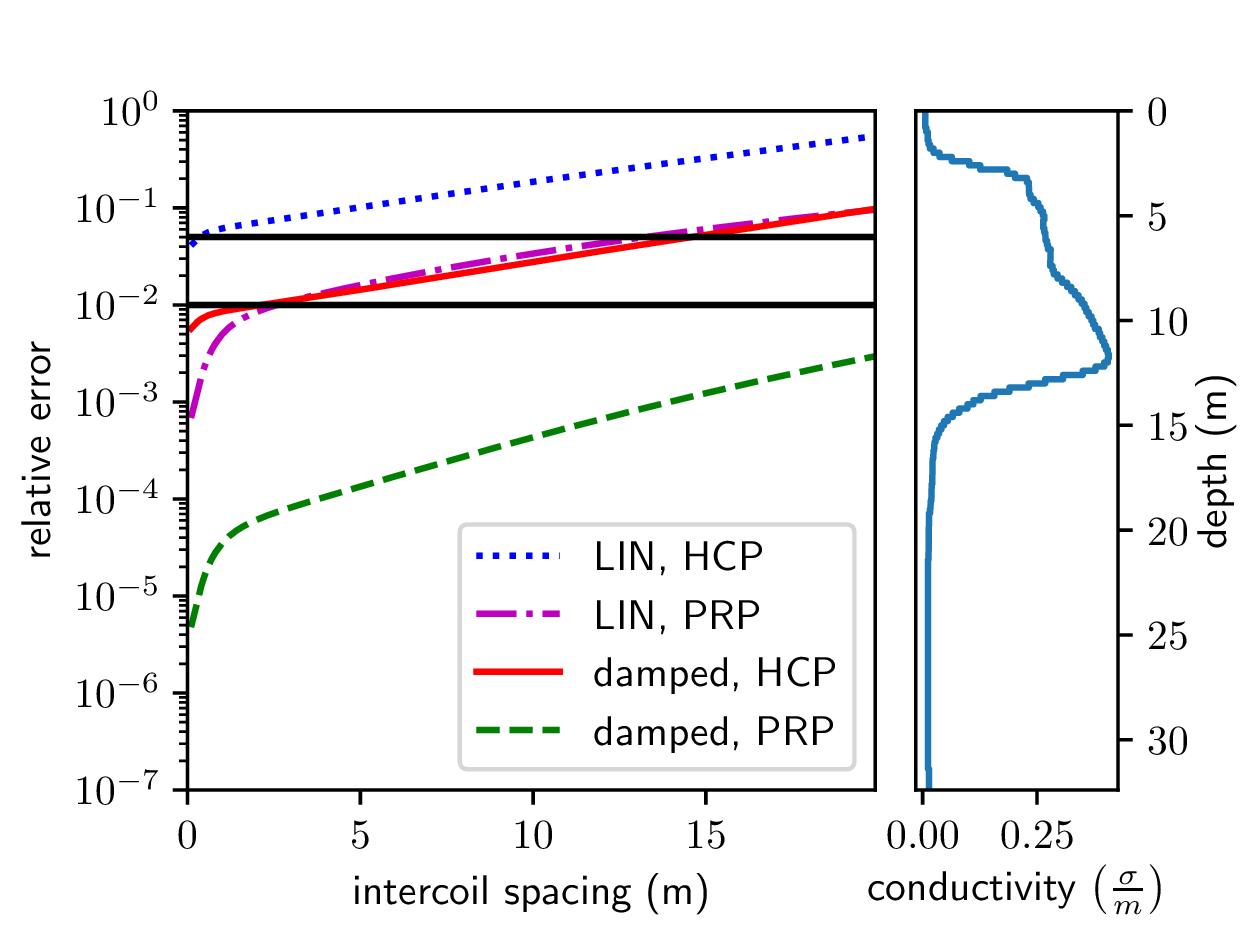}}
  \caption{Relative error of the LIN model and the damped model. The two profiles were obtained from borehole data in~\citep{hermans_imaging_2012}. Using the 5\% induction number (0.31) as limit, one can calculate that the maximum $s$ in \textbf{(a)} is 12~m and in \textbf{(b)} is 9~m. Here, we observe that for this specific model the maximum $s$ is actually larger than the conservative estimate following from the general thumb rule derived in Section~\ref{sec:comp-two-meth}.}\label{fig:saltwater}
\end{figure}
\begin{figure}[t!]
  \centering
  \subfloat[HCP orientation]{%
    \includegraphics[width=.48\textwidth]{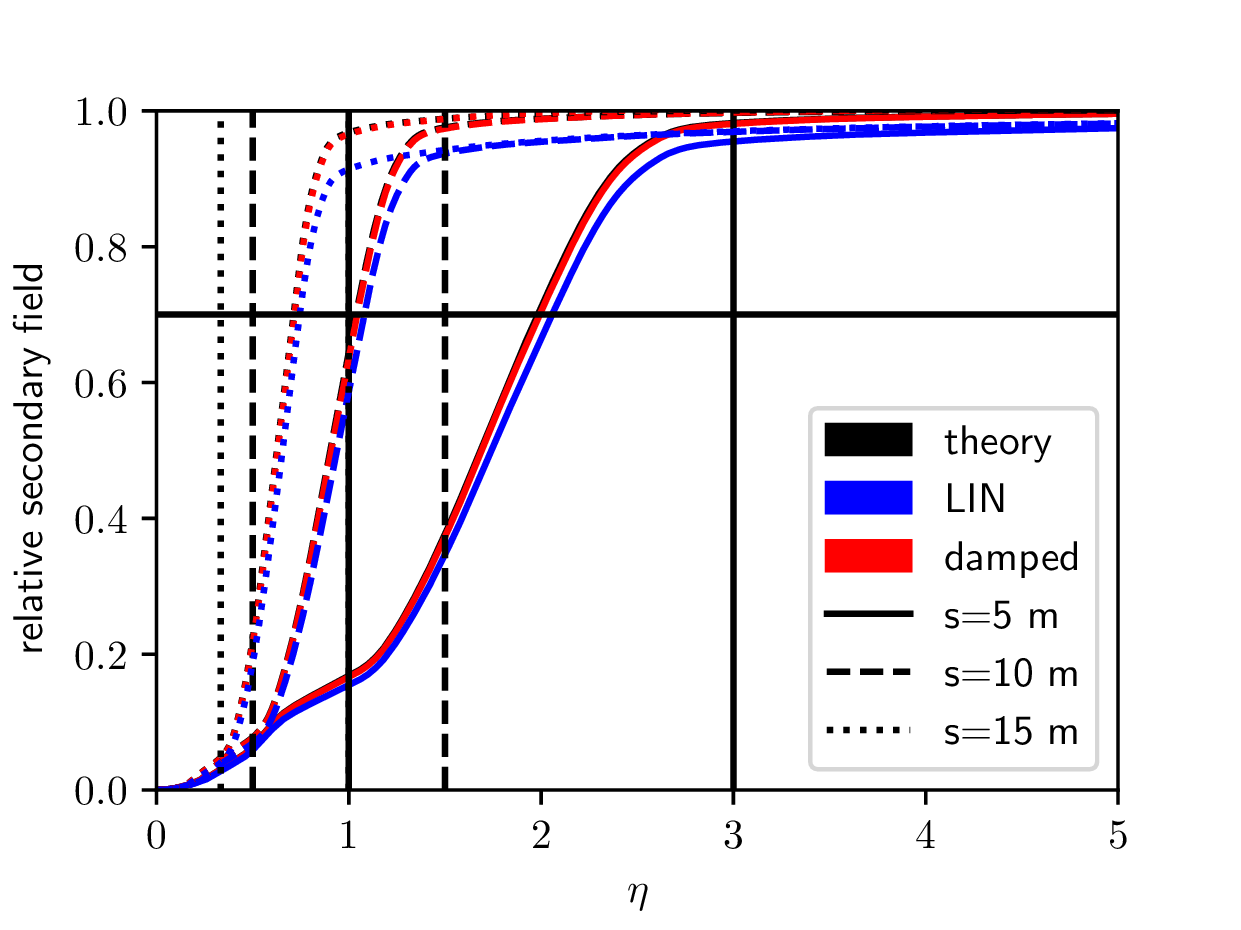}}\hfill%
  \subfloat[PRP orientation]{%
    \includegraphics[width=.48\textwidth]{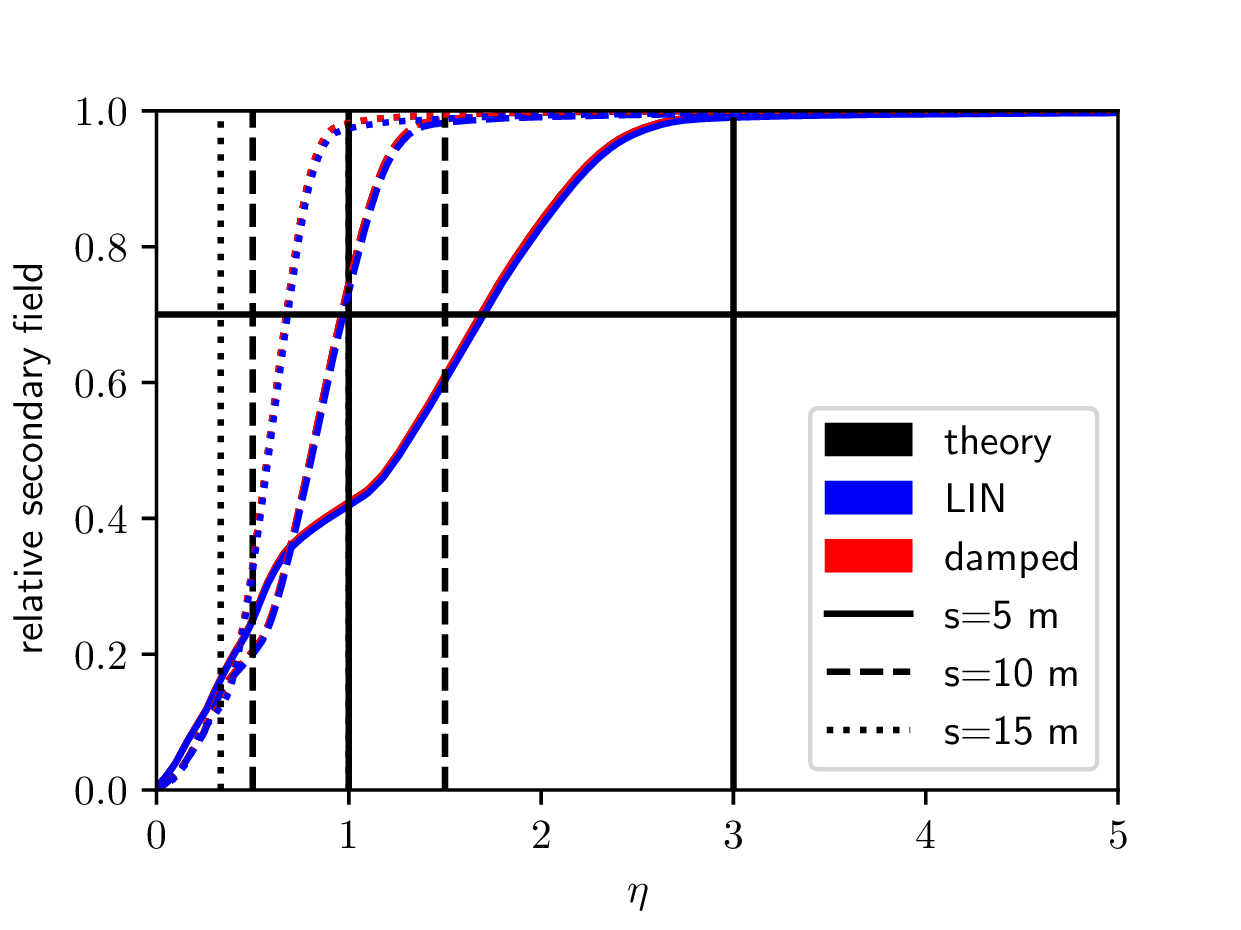}}
  \caption{The sensitivity as function of the normalised depth. The sensitivity is plotted for $s= 5$~m (solid line), $s= 10$~m (dashed line) and $s= 15$~m (dotted line). The horizontal line indicates the 70\% threshold. The vertical lines correspond with the start (5~m) and end (15~m) of the conductivity peak. See Figure~\ref{fig:saltwater}\textbf{(a)} for the conductivity profile.}\label{fig:saltSens}
\end{figure}

As a second, perhaps more stringent, test of our model, we apply it to a conductivity profile inspired by the data gathered in~\cite{hermans_imaging_2012}. It approximately describes the physical situation of a dune area of a Natural Reserve (Westhoek, Belgium). Due to the infiltration of seawater, the conductivity of the soil shows sharp peaks.

In Figure~\ref{fig:saltwater}, the relative error is plotted as function of the intercoil distance. To determine the maximum intercoil distance before the LIN assumption breaks down, we use the induction number (0.31) determined in the previous section. This limit is a sufficient condition to have an error smaller than 5\%. We also remark that the error in the PRP orientation is very small.

Finally we also plot the sensitivity of our model. In Figure~\ref{fig:saltSens} it is plotted for three intercoil distances. Despite the fact that in the PRP orientation the magnetic field is mainly influenced by the upper layers, the effect of the conductivity peak is at least 60\% ($s= 5$~m). As expected the field in the HCP orientation is almost completely determined by the peak (at least 85\% for $s=5$~m).

We can conclude that our model has an error which remains below 5\%, even for the largest intercoil distances. These larger intercoil distances are required to detect the soil deeper in the ground. From these results, we expect that inversion should be able to determine the conductivity peak.

\section{Conclusion}
\label{sec:conclusion}

We have introduced a new model for EMI surveys, summarized in Equations~\eqref{eq:71b}-\eqref{eq:71}, referred to as the damped model, with an error almost ten times smaller than the currently still frequently used  analytical LIN model. If one uses this model, the induction number of the survey should be smaller than 0.029 while for the new model the upper limit is 0.31. Another advantage of our model is the good approximation of the vertical sensitivity, allowing it to be applied for the detection of interfaces between layers. Our model depends on only one additional parameter, a kind of background conductivity, which can be determined from the conductivity and the thickness of the layers. This background simulates the interaction and associated dampening of the electromagnetic fields, between soil layers, a physical effect not present in the LIN model. The resulting equations, only slightly more complicated than the LIN equations, can be straightforwardly implemented. It is important to stress here that the damped model is more convenient to compute with, given its closed-expression format, when compared to the highly nonlinear exact solution which requires an iterative construction of the secondary magnetic field, requiring a numerical integration of an oscillatory integrand. Furthermore if one measures below the aforementioned induction number, the error is negligible in comparison with the measurement error.

Despite being a relatively simple model, it still is reasonably precise, from which we expect an (at least) numerically much more efficient inversion than when using the exact solution or a finite element simulation. The next test of the new model will of course be the setup of an inverse problem.

Finally, although the presented damped model has been developed in frequency space for relatively low frequencies, by Fourier transformation, the construction of a corresponding model in the time domain is also feasible, as long as the situation is such that the underlying frequencies of the transformed dipole current do not get too high. This is also necessary to maintain compatibility with the a priori omitted displacement currents. From this perspective, it is instructive to keep in mind that electromagnetic fields at higher frequencies are even more damped in a conductive setting.

\section{Acknowledgments}
We are grateful to H.~De Gersem, L.~Halleux and in particular T.~Hermans for useful discussions and the providing of experimental data sets. This research did not receive any specific funding and the authors declare no conflicts of interest.

\appendix{}
\section{Horizontal dipole}
\label{sec:horizontal-dipole}

In case of a horizontal dipole the calculations become more cumbersome due to the loss of cylindrical symmetry. This can be circumvented by considering a magnetic monopole instead of a dipole~\citep{wait_geo-electromagnetism_1982}. After calculating the secondary field caused by the monopole we can transform this to the solution in case of a dipole. This can be done using the following operator:
\begin{equation}
  \label{eq:1}
  \left.\frac{\vec{m}}{q} \cdot \vec{\nabla}_{\vec{r}^{\prime}} \right|_{\vec{r}^{\prime}=\vec{0}},
\end{equation}
where $q$ and $\vec{m}$ are respectively the strength of the monopole and dipole. The position of the monopole is $\vec{r}^{\prime}$ while the position of the observer is $\vec{r}$.

After some tedious calculations we get the secondary magnetic field caused by a horizontal dipole $\vec{m}= m\vec{e}_{y}$ at the origin and above a horizontally stratified earth:
\begin{align}
  \label{eq:6}
  H_{s,\prpv} =& \frac{m}{4\pi}\frac{xy}{s^{2}}\int\limits_{0}^{\infty}\lambda^{2} f(\lambda)\exp{(-\lambda z)} \left[\frac{2}{s}J_{1}(\lambda s) - \lambda J_{0}(\lambda s)\right]\d \lambda,\\
  H_{s,\vcp} =& \frac{m}{4\pi}\frac{ 1}{s^{2}}\int\limits_{0}^{\infty}\lambda^{2} f(\lambda)\exp{(-\lambda z)} \left[\frac{y^{2}-x^{2}}{s} J_{1}(\lambda s) - \lambda y^{2}J_{0}(\lambda s)\right]\d \lambda,\\
  H_{s,\prph} =& \frac{m}{4\pi}\frac{y}{s}\int\limits_{0}^{\infty}\lambda^{3} f(\lambda)\exp{(-\lambda z)} J_{1}(\lambda s)\d \lambda.
\end{align}
This magnetic field is measured at a receiver in the same horizontal plane as the dipole, but at an arbitrary position $(x, y)$. There are three possible orientations: perpendicular with a vertical receiver loop (PRP,V), vertical coplanar (VCP) and perpendicular with a horizontal receiver loop (PRP,H). The function $f(\lambda)$ is the same function as Equation~\eqref{eq:12} in Section~\ref{sec:independent-sheets}.

\subsection{LIN model}
\label{sec:mcneill}

Using the same derivation as in Section~\ref{sec:independent-sheets} we obtain the LIN approximation for a horizontal dipole. After integration and normalisation we get:
\begin{align}
  \label{eq:7}
  f(\lambda) &= \frac{k^{2}\d h}{2}\exp{(-2\lambda h)},\\
  h_{s,\prpv}(x,y) &= \frac{-i\mu_{0}\omega}{4}xy \int_{0}^{\infty}2\sigma(\eta s) \left(2 - \frac{4\eta}{\sqrt{4\eta^{2}+1}} -\frac{2\eta}{{\left(4\eta^{2}+1\right)}^{\nicefrac{3}{2}}}\right) \d \eta,\\
  h_{s,\vcp}(x,y) &= \frac{-i\mu_{0}\omega}{4}s^{2} \int_{0}^{\infty}\sigma(\eta s) \left[\frac{y^{2}-x^{2}}{s^{2}}\left(2 - \frac{4\eta}{\sqrt{4\eta^{2}+1}}\right) -\frac{y^{2}}{s^{2}}\frac{4\eta}{{\left(4\eta^{2}+1\right)}^{\nicefrac{3}{2}}} \right] \d \eta, \\
  h_{s,\prph}(x,y) &= \frac{-i\mu_{0}\omega}{4}ys \int_{0}^{\infty}\sigma(\eta s)\frac{2}{{\left(4\eta^{2} + 1\right)}^{\nicefrac{3}{2}}} \d \eta.
\end{align}
In case we measure the secondary field on the $x$-axis ($y=0$) we obtain the result from \cite{mcneill_electromagnetic_1980}.

\subsection{Damped model}
\label{sec:damped-model}

For the damped model the relevant kernel is:
\begin{align}
  \label{eq:8}
  f(\lambda) &\approx  \frac{2\lambda^{2}}{k_{b}^{2}}\frac{\sigma(h)}{\sigma_{b}} {(\lambda-\gamma_{b})}^{2} \exp{(-2\gamma_{b} h - 2\lambda h_{0})}\d h\\
  &\approx \frac{i\mu_{0}\omega\sigma(h)\d h}{2}\exp{(-2\gamma_{b}h)},
\end{align}
where we used the first order expansion for $\gamma_{b}$. For a dipole lying on the ground and a stratified earth, the contribution of the $i$\textsuperscript{th} layer to the secondary field is:
\begin{align}
  \label{eq:9}
  h_{i, \prpv} &\approx \frac{i\omega\mu_{0}\sigma_{i} }{4}xy {\left[ 2I_{\nicefrac{1}{2}}(r_{-})K_{\nicefrac{1}{2}}(r_{+}) - \frac{1}{\sqrt{4\eta^{2} + 1}}\exp{(-ks\sqrt{4\eta^{2} + 1})} \right]}_{\eta_{i}}^{\eta_{i+1}}, \\
  h_{i, \vcp} &\approx \frac{i\omega\mu_{0}\sigma_{i}}{4} s^{2} {\left[ \frac{y^{2}-x^{2}}{s^{2}}I_{\nicefrac{1}{2}}(r_{-})K_{\nicefrac{1}{2}}(r_{+}) - \frac{y^{2}}{s^{2}}\frac{\exp{(-ks\sqrt{4\eta^{2} + 1})}}{\sqrt{4\eta^{2} + 1}} \right]}_{\eta_{i}}^{\eta_{i+1}},\\
  h_{i, \prph} &\approx \frac{-i\omega\mu_{0}\sigma_{i}}{4} s^{2} {\left[\frac{ky}{2\sqrt{4\eta^{2}+1}}\left(I_{1}(r_{-})K_{0}(r_{+}) - I_{0}(r_{-})K_{1}(r_{+})\right)\right]}_{\eta_{i}}^{\eta_{i+1}}.
\end{align}

\end{document}